\documentclass[showpacs, preprintnumbers, nofootinbib, aps, prd, superscriptaddress,10pt, showkeys, notitlepage, twocolumn]{revtex4-1}

\usepackage{graphicx,amssymb,amsmath,amsthm,amsfonts,mathrsfs,epsfig}

\usepackage[linktocpage]{hyperref}
\usepackage[usenames,dvipsnames]{color}
\usepackage{epstopdf}
\usepackage{aas_macros}
\usepackage{pifont}
\definecolor{darkred}{rgb}{0.5,0,0}
\definecolor{darkgreen}{rgb}{0,0.5,0}
\definecolor{darkblue}{rgb}{0,0,0.5}
\definecolor{prussian}{rgb}{0.0, 0.19, 0.33}
\definecolor{richelectricblue}{rgb}{0.03, 0.57, 0.82}
\definecolor{teal}{rgb}{0.0, 0.5, 0.5}
\definecolor{mediumseagreen}{rgb}{0.24, 0.7, 0.44}
\definecolor{lust}{rgb}{0.9, 0.13, 0.13}
\definecolor{ballblue}{rgb}{0.13, 0.67, 0.8}
\definecolor{darkcyan}{rgb}{0.0, 0.55, 0.55}
\definecolor{mountainmeadow}{rgb}{0.19, 0.73, 0.56}
\definecolor{palecarmine}{rgb}{0.69, 0.25, 0.21}
\definecolor{richcarmine}{rgb}{0.84, 0.0, 0.25}
\definecolor{tangelo}{rgb}{0.98, 0.3, 0.0}
\definecolor{venetian}{rgb}{0.784,0.031,0.082}
\definecolor{bdfrance}{rgb}{0.192,0.549,0.906}

\hypersetup{colorlinks=true,
            citecolor=venetian,
            linkcolor=bdfrance,
            urlcolor=lust}

\usepackage{amsmath,amssymb}
\usepackage{tensor}
\usepackage{mathtools}
\usepackage{amsbsy}
\usepackage{bm}
\usepackage{float}

\newcommand{\be}{\begin{equation}}
\newcommand{\ee}{\end{equation}}
\newcommand{\bear}{\begin{eqnarray}}
\newcommand{\eear}{\end{eqnarray}}

\newcommand{\p}{\prime}
\newcommand{\pp}{\prime\prime}

\newcommand{\cL}{{\cal L}}
\newcommand{\cE}{{\cal E}}

\newcommand{\cA}{{\cal A}}

\newcommand{\cQ}{{\cal Q}}

\newcommand{\xhat}{\mathbf{\hat{x}}}
\newcommand{\yhat}{\mathbf{\hat{y}}}
\newcommand{\zhat}{\mathbf{\hat{z}}}

\newcommand{\bB}{\mathbf{B}}
\newcommand{\bv}{\mathbf{v}}
\newcommand{\bfL}{\mathbf{f}_{\rm L}}
\newcommand{\bnabla}{\boldsymbol{\nabla}}

\def\apj{{ApJ}}

\def\apjs{{The Astrophysical Journal Supplement}}
\def\apjl{{ApJL}}

\def\aap{{A\&A}}

\def\mnras{{MNRAS}}

\def\pasj{{Publications of the Astronomical Society of Japan}}

\def\nat{{Nature}}

\def\apss{{Ap\&SS}}
\def\04a{{2004 a}}
\def\04b{{2004 b}}

\begin{document}

\title{Dynamics of magnetoviscous warped discs around compact objects}

\author{Arthur G. Suvorov}
\affiliation{Theoretical Astrophysics, Eberhard Karls University of T{\"u}bingen, T{\"u}bingen, D-72076, Germany}
\affiliation{Departament de F{\'i}sica, Universitat d'Alacant, Ap. Correus 99, E-03080 Alacant, Spain}

\author{Kostas Glampedakis}
\affiliation{Departamento de F\'isica, Universidad de Murcia, Murcia, E-30100, Spain}
\affiliation{Theoretical Astrophysics, Eberhard Karls University of T{\"u}bingen, T{\"u}bingen, D-72076, Germany}

\label{firstpage}

\begin{abstract}

\noindent Accretion discs that are tilted with respect to their compact hosts can warp out-of-plane through general relativistic frame-dragging. 
Warp influences disc dynamics in ways that have been studied extensively, especially as regards instabilities that might lead to rapid angular-momentum cancellation between neighboring rings of fluid and mass infall. 
We provide a review of warped-disc phenomena here, revisiting key hydrodynamical assumptions that impact calculations of the shear viscosity controlling instability thresholds. 
Relativistic effects at the level of gas-parcel orbits are included, as are external Lorentz forces applied by the compact primary's magnetic field. 
Semianalytic analysis reveals that intense magnetic fields can bring about new branches of warp modes and avoided crossings that significantly reduce the perpendicular viscosity 
at sub-Eddington accretion rates. Critical strengths required for misaligned torques to tear a thin disc may thus relax for systems like neutron star X-ray binaries or radio-loud active 
galactic nuclei.
\end{abstract}

\maketitle


\section{Introduction} 
\label{sec:intro}

Accretion discs commonly encircle compact objects in astrophysical environments.
The rate at which energy and angular momentum are transported in these systems depends on a number of factors, such as the orbital dynamics, optical depth, 
and thermodynamic equation of state \cite{fkrbook92}.  Arguably, though, the most critical factor is viscosity: it is ultimately through dissipative processes that matter 
is able to accumulate onto the host's surface or, in the case of a black hole, fall through the horizon.

The benchmark model for accretion dynamics remains, in many respects, that pioneered by \citet{ss73}.  The framework famously comes with an isotropic viscosity parameter, $\alpha$, 
meaning that horizontal and vertical shears are damped at the same rate. However, because of the generally-small aspect ratio of astrophysical discs, this implies that the shear viscosity 
coefficients in orthogonal directions may not only be unequal but differ by orders of magnitude \cite{pp83,og99,lp07}. Quantifying their relative size is central to mapping out onset criteria for magnetohydrodynamic (MHD) or other instabilities, often invoked to explain observed episodes of enhanced accretion \cite{bh98,dub01,nixon15,dogan18,raj21b}.

Due to some combination of general relativistic (GR) effects \cite{bp75}, tidal forces \cite{pap95}, gaseous lifting \cite{mo10}, or irradiation \cite{pringle96}, the disc may 
be off-centred relative to the equator of the primary. Tilt opens new possibilities for instability as, if the host also rotates rapidly, Lense-Thirring (LT) torques act on parcels of disc fluid 
in a directed manner such that they \emph{warp} out-of-plane. Numerical \cite{raj21} and observational \cite{hern96} evidence indicates that, for low enough $\alpha$, 
warp modes may overwhelm restorative viscous forces in the disc. As described by \citet{nix12} and others, it may thus \emph{tear} into a number of precessing rings at certain characteristic 
radii determined by torque balance. As these rings collide and lose angular momentum, large and quasi-periodic mass deposits are made.
To accurately assess whether tearing can explain observational puzzles -- such as the rapid growth of black holes in the early universe \cite{luca25} or outbursts 
in low-mass X-ray binaries (LMXBs) \cite{hein25} -- one should include as many physical ingredients as possible when estimating viscosity coefficients.

In this paper, we add an extra layer of realism to the calculation by accounting for (i) GR adjustments to the epicyclic frequencies of gas parcels, and (ii) magnetic 
fields sourced by the primary.
Aside from providing a critical review of the warped-disc literature -- which can be difficult to navigate owing to swathes of notation and approximation -- we find that 
external Lorentz forces can significantly change the nature of warp eigenmodes (see also Ref.~\cite{po18}). 
This impacts the horizontal-to-vertical viscosity relationship in ways we explore semi-analytically. 
Indeed, while GRMHD simulations have been carried out to examine instability thresholds, with the aim of determining where tearing may 
occur and how the system responds, they cannot reliably explore the whole parameter space due to computational expense (see Ref.~\cite{fl25} for a recent review). 
By using a hybrid analytic-plus-numerical approach, we hope to get a broader overview of how each piece of physics impacts warped-disc dynamics.

This paper is organised as follows. 
The thin-disc structure equations are reviewed in Section~\ref{sec:warptheory}, including how warps may be accounted for (Sec.~\ref{sec:warpgeom}), GR aspects 
related to realignment through the Bardeen-Petterson effect (Sec.~\ref{sec:bp}), and tearing theory (Sec.~\ref{sec:tearing}). A local analysis of how fluid parcels within 
a warped disc may behave is presented in Section~\ref{sec:local}, comparing the Newtonian and GR regimes (Sec.~\ref{sec:dichotomy}), expectations for the viscous 
shear coefficients (Sec.~\ref{sec:alphas}), and hydrodynamical instabilities as a function thereof (Sec.~\ref{sec:hydro}). The main results concerning the inclusion magnetic 
fields are given in Section~\ref{sec:magfield}, where we systematically revisit all previous aspects, notably discussing mode bifurcations and avoided crossing phenomena 
(Sec.~\ref{sec:maglocal}). The impact of GR and magnetic effects on expectations for astrophysical systems is examined in Section~\ref{sec:astro} for a variety of systems, 
including LMXBs (Sec.~\ref{sec:lmxbs}) {and active galactic nucleii (AGN; Sec.~\ref{sec:agn})}. Some closing discussion is offered in Section~\ref{sec:discussion}.

\section{Warped disc phenomenology}
\label{sec:warptheory}

We begin by reviewing some of the basic principles associated with warped disks as found in the literature. More careful analyses at the level of local fluid displacement 
(Sections~\ref{sec:local} and \ref{sec:magfield}), where the physical principles are explored mathematically, are deferred for now 
in favour of a more pedagogical treatment.

\subsection{Basic equations and relations of disc structure} 
\label{sec:structure}

Without warp, the system is assumed to be a standard \citet{ss73} thin disc with thickness $H \ll r$ and viscosity parameter $\alpha < 1$. 
We use standard notation for the disc parameters for coordinate radius $r$: $\Omega (r)$ is the angular velocity profile (typically assumed  to be Keplerian), $\Sigma$ 
is the surface density, $\nu = \eta/\rho$ is the dynamical viscosity coefficient, and $c_s$ is the speed of sound. Useful textbook relations between them are given 
below~\cite{fkrbook92}:
\be \label{eq:diskrelns}
\Sigma = \rho H, \quad  \dot{M} \approx \nu \Sigma, \quad c_s \approx H \Omega = \frac{H}{r}  v_\varphi,
\quad \nu \approx \alpha H c_s,
\ee
where $v_\varphi = r \Omega$ is the rotational (azimuthal) speed. From these expressions we can obtain the density profile of a Keplerian disc
\be
\rho (r) = \frac{\dot{M}}{3\pi \alpha}   \left ( \frac{H}{r} \right )^{-3} (G M r^3 )^{-1/2}.
\label{eq:density1}
\ee 
Unit vectors are denoted by a `hat' while a prime denotes a derivative with respect to the function's argument.
We use $\mathbf{N}$ and $\mathbf{T}$ for the total torque and the torque exerted per unit mass, respectively. 
The disc's angular momentum per unit area is denoted by $ \boldsymbol{\ell}$; for a thin equatorial disc this is
\be
\boldsymbol{\ell} = \Sigma r^2 \Omega\, \mathbf{\hat{z}}. 
\ee
When convenient, we use dimensionless parameters for the radius, $\bar{r} = r c^2/ G M$, and the central body's spin parameter, $q  =  cJ / G M^2$ ($M$ and $J$ 
denote the primary's mass and angular momentum) and/or relativistic units $G=c=1$.  It is also convenient to introduce the Eddington accretion rate
\begin{equation} \label{eq:eddingtonrate}
\dot{M}_{\rm edd} \approx 1.5 \times 10^{-8} (R/10 \text{ km})\ M_{\odot} \text{ yr}^{-1},
\end{equation}
to set units for $\dot{M}$, where $R$ denotes the radius of the compact object (set to the `canonical' neutron star radius with a bit of foresight). 
The structure equations predict a nearly uniform disc thickness ratio which we show here for parameters normalised to those
of an accreting neutron star (and omitting a small relativistic factor):
\begin{align}
 \frac{H}{r}   \approx\, & 2 \times 10^{-3} \left ( \frac{\alpha}{0.1} \right )^{-1/10} \left ( \frac{\dot{M}}{10^{-5} \dot{M}_{\rm edd} } \right )^{3/20}
\left ( \frac{R}{10\,\mbox{km}} \right )^{11/40}
\nonumber \\ 
& \times \left ( \frac{M}{1.4 M_\odot} \right )^{-3/8}   \left (  \frac{r}{R} \right )^{1/8}.
\end{align}
When convenient, we will use the shorthand notation $H_0 = H/r$ for this ratio and keep $H_{0}$ as effectively free to explore a wider range of dynamical interactions.

\subsection{Warped disc geometry} 
\label{sec:warpgeom}

In the case of a warped disc it is useful to introduce the unit tilt vector,  $\hat{\boldsymbol{ \ell}} =  \boldsymbol{\ell} / \ell$. 
By its definition this is orthogonal to the local disc/orbital plane (e.g. for the standard flat thin disc model we would have $ \hat{\boldsymbol{ \ell }} = \mathbf{ \hat{z}} = \mbox{const.} $).
In general, this vector is a function  $\hat{\boldsymbol{ \ell}} (t, \mathbf{x} )$ of the fixed global coordinates. However, we can use `warp-adapted' 
coordinates \cite{pp83} (henceforth PP83) so that each disc ring at distance $r$ from the center is axisymmetric and, as a result,  
$\hat{\boldsymbol{ \ell}} =\hat{\boldsymbol{ \ell}} (t, r )$, where $r$  can be a spherical or a cylindrical radius. These adapted coordinates emerge `naturally' by considering 
the local exchange of angular momentum between adjacent rings in the warped disc \cite{og99}.

With the help of the tilt vector, we can define the vectorial warp as
\be \label{eq:vectorwarp}
\boldsymbol{\psi} (t, r) = r  \frac{ \partial   \boldsymbol{\hat{\ell}} }{\partial r}. 
\ee
Roughly speaking, the warp parameter encodes the radial variability of the disc's tilt. Throughout most of this work we use a `small-warp' approximation, amounting 
to $|\psi| \ll 1$, to explore perturbative corrections to the disc structure equations; see, however, \citet{og99} for a discussion on the validity of this approach and its 
comparison with the full Navier-Stokes system for warped discs.

\subsection{Bardeen-Petterson alignment} 
\label{sec:bp}

One aspect of relevance to warp considerations is the \citet{bp75} (BP) effect: it has been argued that the inner region of a disk should gradually align with 
the equatorial plane of the compact primary (though cf. Refs.~\cite{kp85,sch96}). The effect stems from the assumption that perturbations between neighbouring rings of 
gas, in a tilted disc subject to radially-dependent  (i.e. LT) torques,  propagate in a diffusive regime. Such diffusive processes cause any angular-momentum 
misalignments to be smoothed out to some transition radius ($r_{\rm BP}$) set by the competition between viscosity and differential precession. Beyond this critical 
radius, LT torques become small and tilt is set instead by the direction of gas infall, leading to the picture of  a midplane-oriented disc with thickened `arms' extending at 
some angle; see, e.g., Fig. 1 in \citet{frag01}. If the BP effect operates, one would not expect any significant warp in the inner  regions of the disc owing to its alignment 
and hence the system may be in a (quasi-)equilibrium state.

Such an application is subtle however. PP83 subsequently argued that even if the disc viscosity is assumed isotropic, damping timescales in the vertical and radial 
directions are not equal owing to the small aspect ratio of the disc (such a relationship is quantified in Sec.~\ref{sec:local}). This, together with the fact that disturbances 
may instead propagate in a wave-like  regime for some microphysical combinations (see below), implies that the BP effect may not occur and instead more catastrophic 
phenomena in the form of \emph{tearing} can take place.

\subsection{Fragmentation and tearing} 
\label{sec:tearing}

\begin{figure*}
\centering
 \includegraphics[width=0.7\textwidth]{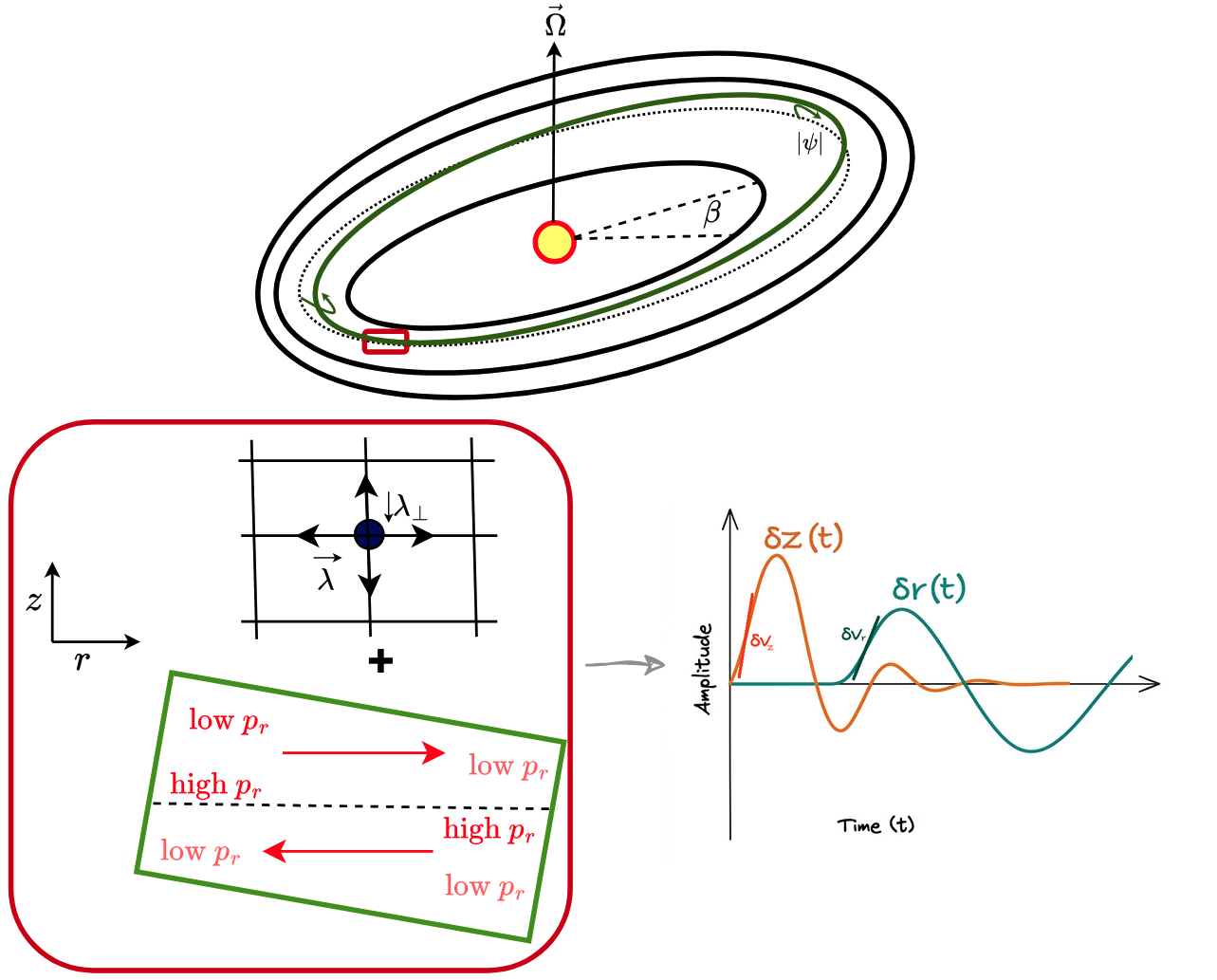}
 \caption{{Schematic of a tilted accretion disc around a compact object and the dynamics of local fluid parcels within. In cases of dynamical capture, young systems, 
 or where the BP effect is not at work, there is no further expectation of any existing disc-object symmetry: the inner region(s) may be angled ($\beta$) with respect to the mid-plane 
 of the primary due to Lense-Thirring torques or otherwise. A differential precession is thus exerted on radially-neighbouring parcels of fluid which drives a warp ($|\psi|$).
Within a small parcel oriented with respect to locally Cartesian coordinates (red inset), two effects operate: (1) because of viscosity in the vertical ($\lambda_{\perp}$) and radial ($\lambda$) directions, oscillations are inherently damped. (2) due to radial pressure gradients formed between neighbouring `rings', which vary as a function of polar angles, oscillations are coupled because the disk pressure varies with height.
The result is that a driving warp perturbation issues a radial response, as shown by bottom right panel, the former of which is damped more rapidly than the latter if $\lambda_{\perp} > \lambda$: balancing the energy-dissipation rates between shear motions in two orthogonal directions leads to estimations of these quantities (see text).
The radial and vertical velocities are given by the slopes of the respective amplitudes of oscillation.
(Parts of this figure were adapted from those presented in sections 2 in \protect\citet{dulle22} and 4.1 in \protect\citet{lp07}).}}
  \label{fig:schematic}
\end{figure*}

Although a mixed scheme is likely to apply in reality, the perturbation dynamics of a warped disc are typically discussed in two distinct regimes. These correspond to 
either $\alpha \ll H/r$, where viscous damping is weak enough for modes to effectively propagate in a wavelike manner, or $\alpha \gg H/r$, where modes may be damped 
before completing full oscillations (a diffusive regime where the BP effect may operate; see Refs.~\cite{pringle92,nix12}). It is typically only in the wavelike case that stress may 
build up in such a way that a disc may be locally overwhelmed and `break' in some sense. 

Following \citet{raj21}, the word \emph{break} used above simply conveys that there exists a mode, scaling as $\propto e^{i \omega t}$, such that the complex eigenfrequency 
$\omega$ has a negative imaginary component. This implies a runaway instability where the mode amplitude grows, meaning that the perturbed warp, $\delta \psi$, may become 
large enough to snap the disc at some position. The word \emph{tearing}, by contrast, is typically applied in the context where external (i.e., primary or companion sourced) 
torques act on the disc and are necessary for inducing such a break. They convey the same physics therefore, though it is the latter which is of primary relevance since an
 intrinsically-unstable disc is unlikely to persist over astrophysical timescales. In the case of primary-sourced rotation, \citet{nk12} argue that viscous torques cannot override 
 the LT torque if the warp exceeds a critical amplitude, leading to fragmentations into precessing rings within the disc mid-plane. Precessional grinding between these rings 
 leads to the eventual loss of their angular momentum such that the gas then falls quasi-radially onto the host object, increasing $\dot{M}$ relative to a quiescent state \cite{nix12}. 
 An illustration of how warp and precessional grinding may manifest within a tilted disc is provided in Figure~\ref{fig:schematic}.

Either way, such an analysis -- revisited via a hydrodynamic description in Sec.~\ref{sec:hydro} -- only indicates the physical parameters under which 
a break may occur. It does not, however, identify the radial position, which is important to quantify if one wishes to estimate the amount of mass that may be contained 
in a broken region that may eventually contribute to a brightening episode (see Sec.~\ref{sec:astro}).  

To estimate the tearing radius, $r_{t}$, we work with torques per unit mass and closely follow \citet{td14}. To leading post-Newtonian order, the LT torque formula is
\be
\mathbf{T}_{\rm LT} = \frac{2 c^2 q}{\bar{r}^{5/2}} \left (   \mathbf{\hat{J}} \times \boldsymbol{\hat{\ell}} \right )
~ \Rightarrow ~ T_{\rm LT} = \frac{2 c^2 q}{\bar{r}^{5/2}} \sin \beta,
\ee
where we have introduced the tilt angle $\beta$. Note that this torque is always orthogonal to the central body's spin axis, $\mathbf{T}_{\rm LT} \cdot \mathbf{\hat{z}} =0$. 
The tearing action of $\mathbf{T}_{\rm LT} $ is counteracted by viscous forces arising from the shearing motion in a direction orthogonal to the disc plane.
As a consequence, we need to introduce a `perpendicular'  viscosity coefficient $\nu_\perp$ which may be distinct from the more familiar `equatorial' viscosity coefficient $\nu$ 
of the standard $\alpha$-disc\footnote{This point becomes aparent if we recall that the viscous torque in a thin Keplerian disc is 
$ \mathbf{N}_{\rm visc} (r) \approx 2 \pi \nu \Sigma r^3 \Omega^\p \mathbf{\hat{z}}$~\cite{fkrbook92}, which is orthogonal to the LT torque.}. 
Using a similar $\alpha$-parametrisation for this perpendicular viscous torque to the standard \citet{ss73} one, we have
\be
T_\perp \sim | \psi | \nu_\perp \Omega  \sim | \psi | \alpha_\perp H^2 \Omega^2 \approx | \psi |\alpha_\perp\left ( \frac{H}{r} \right )^2 \frac{c^2}{\bar{r}}.
\ee
Then, the criterion for disc tearing becomes 
\be
\frac{T_{\rm LT}}{T_\perp} \gtrsim 1 ~ \Rightarrow ~ \bar{r}_t \lesssim \left ( \sin \beta \frac{q}{\alpha_\perp | \psi | } \right )^{2/3} \left ( \frac{H}{r} \right )^{-4/3}.
 \label{eq:tearing}
\ee
This result displays the expected dependence on the tilt angle $\beta$ and the central body's spin $q$. However, the viscosity dependence is a more
subtle issue because of the `anomalous' relation $\alpha_\perp (\alpha)$ which may cause $r_t$ to \emph{grow} with $\alpha$~\cite{pp83, kp85}. 
In fact, one of the main goals of this work to revisit this relationship and see how it changes in the presence of GR effects and central 
body-hosted magnetic fields.

It is important to recognise that the radii $r_{\rm BP}$ and $r_{\rm t}$ essentially represent the same effect in different parts of the parameter space. 
For example, the former radius can describe the case of an aligned disc following the suppression of the LT effect by viscosity (assuming the tilt radius 
was small to begin with). The latter radius would represent the case where the LT force dominates over viscosity, leading to the formation of decoupled 
precessing rings. This dichotomy is exemplified in Fig.~11 of \citet{neal15}: below (above) a critical tilt angle the disc BP-aligns  (fragments into rings) 
under the competitive actions of viscosity and LT precession.

\section{Local fluid-parcel analysis}
\label{sec:local}

\subsection{Equations of motion and their solutions}

As mentioned earlier, a local analysis provides a relatively simple approach to modelling the dynamics of a warped disc with the possibility of
including additional degrees of freedom like relativistic effects and an external magnetic field. 

Following \citet{lp07} (hereafter LP07), we model a disc fluid element as a test body of mass $m$ subject to relativistic gravity
and a viscous drag force.
{This drag should be thought of simply as the resistance the fluid provides in response to oscillations of embedded particles.}
 In addition we include a force induced by the disc warping; a local non-zero warp leads to a radial (locally equatorial)  pressure gradient 
with magnitude (see Fig.~\ref{fig:schematic})
\be \label{eq:partialp}
\partial_r p \sim \partial_z p \, \psi \sim \frac{p}{H} \psi,
\ee
where it is assumed that the warp angle $\psi \sim dz/dr$ is small. The (radial) acceleration associated with this force is, {from equations \eqref{eq:diskrelns}},
\be \label{eq:pressuref}
\frac{1}{\rho} \partial_r p \sim \frac{c_s^2}{H} \psi \sim H \Omega^2 \psi.
\ee
{Numerical prefactors of order a few could appear in the above due to our approximate handling of the derivatives.}
For the viscous force we use the simple Stokes formula, allowing for an anisotropic drag coefficient (see discussion below on this point)
\be \label{eq:dragf}
\mathbf{f}_{\rm visc} = - m \lambda_i \bv \qquad (i = \{x,y,z\}).
\ee
The validity of this formula hinges on a laminar and subsonic flow, which need not be satisfied throughout the disc. Nevertheless, it provides a reasonable approximation for our 
purposes and is standard in studies of linear stability; see Section 4 of \citet{bran08} for a discussion on different regimes. 

The above forces are of hydrodynamical origin and are supposed to act as perturbations to the body's geodesic motion. For the 
latter motion we can use exact equations if the central body is a Kerr black hole or a slow-rotation limit of the same equations if the central body
is a neutron star or some other material system. These textbook equations are shown in Appendix~\ref{app:testbody}. We apply them to the
case of a fluid element perturbed away from a circular equatorial orbit, i.e.
\be
r(t) = r_0 + \delta r (t), \qquad \theta (t) = \frac{\pi}{2} + \delta \theta (t),
\ee
where $\{t,r,\theta\}$  now stand for the standard Boyer-Lindquist coordinates of the Kerr spacetime (ignoring the azimuth). As shown in Appendix~\ref{app:testbody},
the motion is described by the two decoupled harmonic oscillators in the absence of hydrodynamical forcing
\be
\frac{d^2 \delta r}{dt^2} + \Omega_r^2 \delta r= 0, \qquad   \frac{ d^2 \delta \theta}{dt^2} + \Omega_\theta^2 \delta \theta= 0,
\label{oscillatorsK}
\ee
where we henceforth use an overhead dot to denote (coordinate) time differentiation.
Adopting for the moment geometric units $G=c=1$, we define the dimensionless parameters $q =a/M$ and $ x =  M/r_0 = ( M \Omega_0 )^{2/3} $, 
where $\Omega_0 = \Omega (r_0)$. We can subsequently write the epicyclic orbital frequencies $\Omega_r, \Omega_\theta$ in terms of these parameters as
\begin{align} 
\label{eq:omegarelns}
\Omega_r^2 &= \frac{x^3 - 6 x^4 + 8 q x^{9/2} - 3 q^2 x^5}{M^2(1+q x^{3/2} )^2},
\\ \nonumber \\
\Omega_\theta^2 &= \frac{x^3 - 4 q x^{9/2} + 3 q^2 x^5}{M^2(1+q x^{3/2} )^2}.
\end{align}
As mentioned, these expressions are `exact' for orbits outside black holes but they need to be truncated to ${\cal O} (q)$
for accreting neutron stars as their exterior spacetime is not Kerr (see, e.g., Ref.~\cite{pap17}). In fact, if we want to amend~\eqref{oscillatorsK} with the aforementioned hydrodynamical forces (i.e., expressions \ref{eq:pressuref} and \ref{eq:dragf}) we should
interpret them as quasi-Newtonian equations [the error involved in doing this approximation is ${\cal O} (x)$ in the solutions $\delta r, \delta \theta$].  
The resulting system reads
\begin{align}
\delta \ddot{r} +  \Omega_r^2   \delta r +  \lambda  \delta \dot{r} &=   H \Omega^2_0 \psi (t),
\label{equrad1}
\\   \nonumber 
\\
\ddot{\psi} +  \Omega_\theta^2 \psi  + \lambda_\perp    \dot{\psi} &= 0,
\label{equz1}
 \end{align}
 where we have replaced $\delta \theta \sim \psi$ and introduced the horizontal and perpendicular drag coefficients $ \lambda,  \lambda_\perp$.
{This replacement is motivated physically by the definition \eqref{eq:vectorwarp}: warp measures relative tilts and thus out-of-plane, angular oscillations should set the characteristic scales of local variability.}
Note that the system of equations \eqref{equrad1} and \eqref{equz1} generalise the analysis of LP07 by including relativistic orbital frequencies and a dissipative term in the warp's equation of motion (see their Section 4.1). 
 
 The solution of the warp equation \eqref{equz1} (with an appropriate choice of initial conditions) can be written as
  \be
 \psi (t) = \psi_0  e^{-\lambda_\perp t/ 2}  \sin  \omega_\theta t,
 \ee
 where $\psi_0$ is a constant amplitude and $ \omega_\theta = ( \Omega_\theta^2 - \lambda^2_\perp/4 )^{1/2}$. In the case of weak damping, $\lambda_\perp \ll \Omega_\theta$, 
 we can approximate $\omega_\theta \approx  \Omega_\theta \approx \Omega_0$, and the solution is oscillatory over an interval $t \sim 1/\lambda_\perp$. 
 The opposite case of strong damping, $\lambda_\perp \gg \Omega_\theta$, is characterised by an imaginary 
 $\omega_\theta = i  ( \lambda^2_\perp/4 - \Omega_\theta^2 )^{1/2} = i \tilde{\omega}_\theta$ and the time evolution is a purely exponential decay with a late time 
 profile $\psi \sim e^{(\tilde{\omega}_\theta-\lambda_\perp/2 ) t}$.  These two limiting cases provide a link to the previous torque-based tearing radius calculation:
 weak (strong) damping is associated with a dominant LT (viscous) torque.  A necessary condition for the disc to be driven to a tearing regime 
 is that it should be allowed to oscillate perpendicularly (LT precession) over several orbital periods so that adjacent rings can develop a sufficiently large
 relative displacement. 
 
 Going back to the fluid parcel's radial equation of motion \eqref{equrad1}, we have the updated expression
 \be
 \delta \ddot{r} +  \Omega_r^2   \delta r +  \lambda  \delta \dot{r} =   H \Omega^2_0  \psi_0  e^{-\lambda_\perp t/ 2}  \sin  \omega_\theta t.
 \ee 
 Considering only the particular, warp-driven, solution of this equation, we find
 \begin{align}
 \delta r (t) 
 & =  \cA e^{-\lambda_\perp t/ 2}  \sin ( \omega_\theta t + \phi ), 
 \end{align}
 where $\phi$ is a constant phase and 
 \be 
 \label{eq:hydroamp}
 \cA = \frac{  H   \psi_0 \Omega^2_0 }{ \sqrt{ \lambda_\perp^2 \Omega_r^2 + \lambda^2 \Omega_\theta^2 
 -\lambda \lambda_\perp (\Omega_r^2 + \Omega_\theta^2)  + (\Omega_r^2 - \Omega_\theta^2)^2} }.
 \ee
 The expression for the amplitude can be simplified by approximating $\Omega_{r,\theta} \approx \Omega_0$ in the \emph{viscosity} terms, 
 \be
 \cA \approx  \frac{  H   \psi_0 \Omega^2_0 }{ \sqrt{ ( \lambda_\perp - \lambda )^2  \Omega_0^2   + (\Omega_r^2 - \Omega_\theta^2)^2} }.
 \ee
 We can employ a standard $\alpha$-parametrisation for the damping rates, $\lambda = \alpha \Omega_0$ and $\lambda_\perp = \alpha_\perp \Omega_0$,
to furthermore write 
 \be \label{eq:ampxx}
 \cA  \approx  \frac{  H   \psi_0}{ \sqrt{ ( \alpha_\perp - \alpha )^2  +  \left [ \,(\Omega_r^2 - \Omega_\theta^2)/  \Omega_0^2  \,\right ]^2} }.
 \ee
 {As mentioned earlier, this amplitude may be a factor few larger in reality if we were to use an exact equation of state to write out the pressure derivatives in equations \eqref{eq:partialp}.}
 The structure of this amplitude is quite revealing: an orbital resonance between the two epicyclic frequencies maximises the fluid parcel's radial displacement.
 The resonance is exact in the case of Newtonian gravity ($\Omega_r = \Omega_\theta$) but there is a small shift when GR corrections are included.
 We discuss these two regimes in the following section.

\subsection{Newtonian versus relativistic regime}
\label{sec:dichotomy}

If we assume (as in LP07) a Newtonian gravitational field, so that $\Omega_r = \Omega_\theta = \Omega_0$ exactly, we find the following Newtonian 
oscillation amplitude
\be \label{eq:newtampformula}
\cA \approx \cA_{\rm N} = \frac{H \psi_0}{|\alpha_\perp-\alpha|}.
\ee
Having at our disposal the fully relativistic orbital frequencies~\eqref{eq:omegarelns} we can refine this Newtonian result by first expanding in the 
post-Newtonian parameter $x \ll 1$
\begin{align}
\Omega_r^2 -  \Omega_\theta^2 & = \frac{6x^4}{M^2} \left ( \frac{1 - q x^{1/2}}{1 + q x^{3/2} } \right )^2  
 \nonumber \\
&\approx \frac{1}{M^2} \left [ 6 x^4 - 12 q x^{9/2} + {\cal O} (q^2,x^5) \right ].  \label{eq:omegaexp}
\end{align}
We can then see that the above Newtonian approximation amounts to the condition 
\be
| \alpha_\perp - \alpha | \gg   x.
\label{Nregime}
\ee
(Normal human units can be restored with $x \to GM/ r_0 c^2$.)
In the same Newtonian regime, the radial velocity is
\be
\delta v_r = \delta \dot{r}  \approx \frac{H \Omega_0 \psi_0}{|\alpha_\perp-\alpha|} \approx  \frac{c_s \psi_0}{ |\alpha_\perp-\alpha|}. 
\ee
In the opposite limit, which we may call `relativistic', we have 
\be
 | \alpha_\perp -\alpha |  \ll  x,
 \label{GRregime}
\ee
leading to a viscosity-independent radial motion amplitude
\be
\cA \approx \cA_{\rm GR} =    \frac{H \psi_0}{6 x}.
\ee
The corresponding velocity amplitude is  
\be
\delta v_{r} \approx  \frac{c_s \psi_0}{6 x} .
\ee 
We can see that 
\be
\frac{A_{\rm GR} }{ A_{\rm N}} \approx \frac{|\alpha_\perp -\alpha|}{x},
\ee
which means that the relativistic regime implies a suppressed radial displacement and velocity. This is not surprising given this regime's 
weaker resonance between the radial and vertical oscillation frequencies. 

It is clear that the Newtonian regime is the suitable one for the greatest portion of an astrophysical accretion disc for `canonical' values $\alpha \approx 0.1$ 
(excluding the fine-tuned situation $\alpha_\perp \approx \alpha$). The disc may extend down to the relativistic innermost stable orbit 
(at $x = 1/6 \approx 0.17$ for a non-spinning black hole) which means that the relativistic regime requires the alpha parameters be of order $\sim 10^{-2}$ or less. 
As we show in the following section the refined threshold for the relativistic regime turns out to be $\alpha \ll x^3 $. 
There is, of course, nothing sacred about `canonical' values $\alpha \approx 0.1$. Indeed, as we discuss in Section~\ref{sec:alphavalues}, the anticipated values of
$\alpha$ in astrophysical systems may lie much below (or even much above!) the canonical value, suggesting that the relativistic regime could be physically
relevant in some cases.


\subsection{Relating the two viscosities}
\label{sec:alphas}

As discussed in LP07 {and Sec.~\ref{sec:intro}}, the fact that the horizontal and vertical motions are interrelated (i.e. the horizontal shearing 
motion is driven by the disc's warp) implies a balance in the corresponding viscous dissipation rates. In terms of the fluid velocities in cylindrical coordinates 
$(r, z)$, we should have [see Eq.~(40) and surrounding discussion in LP07]
\be
\nu_\perp \left \langle  \left ( \partial_r \delta v_z  \right )^2 \right  \rangle = \nu \left \langle \left ( \partial_z \delta v_r \right )^2 \right \rangle.
\label{dissbalance}
\ee 
The angled brackets denote an averaging over many hydrodynamical timescales to wash out oscillatory motions. The vertical velocity is
calculated as
\be
\delta v_z \approx r_0 \dot{\psi} \approx  r_0  \Omega_0 \psi_0  e^{-\lambda_\perp t/ 2}  \cos \Omega_0 t,
\ee
and its $r$-derivative can be approximated as $\partial_r \delta v_z  \approx \delta v_z /r_0$.  
For the $z$-derivative of the radial velocity we use the approximation\footnote{This, as well as the previous approximation for $\partial_r$ concerning expression \eqref{eq:pressuref}, 
is oblivious to the sign of the actual derivative; this is not a problem in this instance because we are considering the squared quantities. 
In later sections it will be important to keep track of the sign.}
\be
\partial_z \delta v_r \approx \frac{\delta v_r}{H} \approx \frac{\cA \Omega_0 }{H} e^{-\lambda_\perp t/2} \cos (\Omega_0 t + \phi).
\ee
The time average of these derivatives in the weak viscosity regime ($\lambda_\perp \ll \Omega_0$) is found to be
\begin{align}
 \left \langle  ( \partial_r \delta v_z )^2 \right  \rangle & \approx \frac{1}{2} \psi_0^2 \Omega_0^2,
\\ 
\left \langle  (\partial_z \delta v_r)^2  \right \rangle & \approx \frac{1}{2}  \frac{\cA^2 \Omega_0^2}{H^2}.
\end{align}
Inserting these in the balance equation~\eqref{dissbalance} we obtain
\be 
\frac{\nu_\perp}{\nu} = \frac{\alpha_\perp}{\alpha} \approx    \left ( \frac{\cA}{ \psi_0 H} \right )^2.
\label{perpviscnewt}
\ee
Evaluating this in the Newtonian regime, $ \cA = \cA_{\rm N}$, via expression~\eqref{eq:newtampformula},
\be 
\frac{\alpha_\perp}{\alpha} \approx  \frac{1}{(\alpha_\perp -\alpha)^2} ~\Rightarrow ~ \alpha_\perp \approx  \alpha^{1/3},
\label{alphaperpN}
\ee
where we have kept the single real root of the cubic equation and assumed $\alpha \ll 1$. This solution is consistent with the
condition~\eqref{Nregime} provided $\alpha$ is bracketed as
\be
x^3 \ll \alpha \ll 1.
\label{bracketalphaN}
\ee
In the relativistic regime, $\cA = \cA_{\rm GR}$, we instead find
\be \label{alphaperpGR}
\alpha_\perp \approx  \frac{\alpha}{x^2}.
\ee
together with the constraint $\alpha \ll x^3$ (which also guarantees that $\alpha_\perp \ll 1$). 

Our main result~\eqref{alphaperpN} for the perpendicular viscosity is at odds with the result obtained in LP07. This is due to our inclusion of 
a damping term in the warp's equation of motion~\eqref{equz1}; the removal of that term (which to us appears unjustified) is equivalent to taking 
the limit $|\alpha_\perp - \alpha | \to \alpha$ in the formulae of the last three sections. In particular,~\eqref{alphaperpN} would be replaced by 
$\alpha_\perp/\alpha \approx 1/\alpha^2$ thus leading to the LP07 result $\alpha_\perp \approx 1/\alpha$. Putting aside the  
different functional form $\alpha_\perp (\alpha)$, it is clear that both models predict a dominant perpendicular viscosity,  $\alpha_\perp \approx 0.5 -10$
for the canonical value $\alpha \approx 0.1$.

\begin{figure}
\centering
 \includegraphics[width=0.48\textwidth]{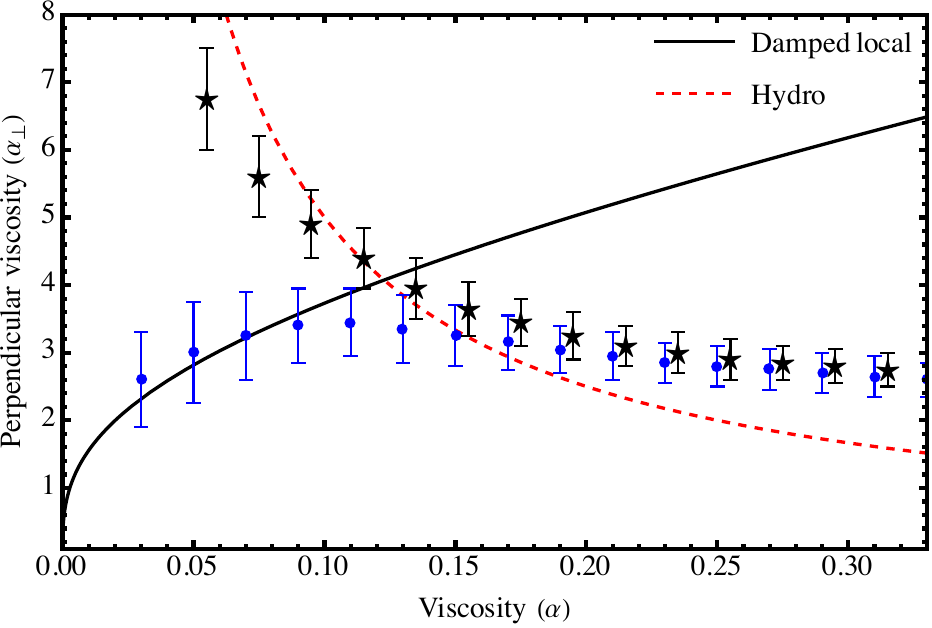}
 \caption{Comparison between relation \eqref{alphaperpN} scaled by a factor 7 and $\alpha_{\perp} = 1/2\alpha$ as a function of SS73 viscosity $\alpha$. Overlaid are data extracted from the simulations of \protect\citet{lod10} (see their figure 6) in cases without (blue circles) and with (black stars) bulk viscosity.}
 \label{fig:differentalpha}
\end{figure}

{The functional form for the modified $\alpha_{\perp}(\alpha)$ relation -- scaled by a factor 7 for demonstration purposes (see discussion below equation \eqref{eq:ampxx}) --  is depicted 
in Fig.~\ref{fig:differentalpha} alongisde the hydrodynamical one ($\alpha_{\perp} \approx 1/2\alpha$) \cite{pp83}.
We overlay data from the perpendicular viscosity inferred from the Newtonian, smoothed-particle-hydrodynamics (SPH) simulations of Ref.~\cite{lod10} (their figure 6 in particular) 
in the case where they include (black stars) or switch off (blue circles) bulk viscosity.
We see that the overturn pattern obtained from relation \eqref{alphaperpN} matches well for $\alpha \lesssim 0.1$ to the simulations when bulk viscosity is absent, as in our local analysis. 
We caution the reader however that \citet{lod10} argue SPH simulations without bulk viscosity can introduce numerical artifacts.
This is because the conservative particle momenta lack a dissipative mechanism to resolve steep gradients, resulting in unphysical interpenetration effects and high-frequency acoustic ringing across shock fronts.
It could be thus simply be a coincidence that the modified relation better fits these data.
Nevertheless, the clear match with the modified relation is curious and hints towards its applicability to certain low-viscosity environments.
}

\begin{figure}
\centering
 \includegraphics[width=0.48\textwidth]{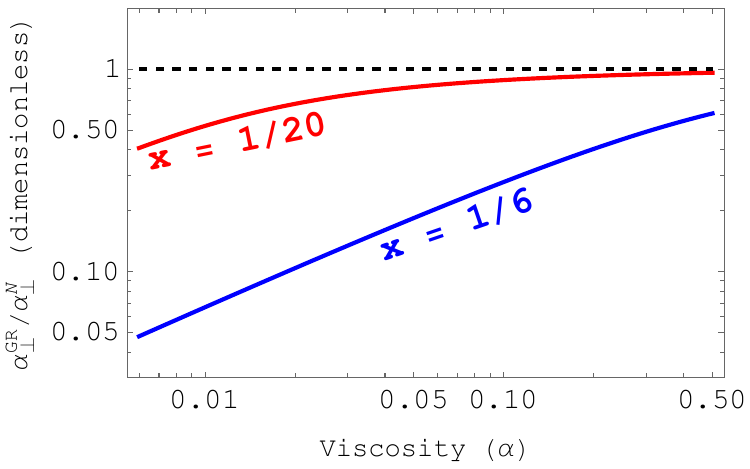}
 \caption{Ratio of the perpendicular viscosity coefficient, determined using the amplitude \eqref{eq:hydroamp} and expression \eqref{perpviscnewt}, to the 
 Newtonian prediction \eqref{alphaperpN} as a function of $\alpha$ for $q=0.2$ and $x=1/6$ (blue) or $x=1/20$ (red). 
 The horizontal line marks equality, reached after $\alpha \approx 0.1$ for the less compact case. Solutions asymptote to this value for large $\alpha$.}
 \label{fig:hydro_viscosities}
\end{figure}

The relative ratio between the full prediction \eqref{eq:hydroamp} and the Newtonian result, expression \eqref{alphaperpN}, 
is shown in Fig.~\ref{fig:hydro_viscosities} for a compact object at $x=1/6$ and $q=0.2$ (blue) or a less-compact system with $x=1/20$ (red). 
Note that such a $q$ value corresponds roughly to a spin frequency $\nu \sim 400$~Hz for a neutron star. 
At low values of the viscosity, we see significant departures: the ratio 
$\alpha_{\perp}/\alpha$ is \emph{lower} by a factor of $\sim 10$ for $\alpha \lesssim 10^{-2}$ for $x=1/6$ and even still $\sim 2$ for $x = 1/20$. 
This implies that, generally speaking, we expect warped discs to be \emph{less stable} when accounting for GR effects, as we explore next.

\subsection{Hydrodynamical estimates for instability} 
\label{sec:hydro}

Two effective viscosities, related to the drag coefficients $\lambda$ and  $\lambda_\perp$, have been introduced above. The formal analysis of \citet{og99} 
reveals a third ought to be considered also, which includes contributions from radial stresses and encapsulates dispersive, wave-like torques which cause 
neighbouring rings of fluid to precess when misaligned (see equation 83 in Ref.~\cite{po18} for a definition).
In this section, we examine how the adjusted amplitude \eqref{eq:hydroamp} affects the stability of a disc using the 
hydrodynamic formalism introduced by \citet{og99} and \citet{dogan18}. 

Adopting the notation of \citet{dogan18}, the relevant hydrodynamical equations in an isothermal disc are that describing 
conservation of mass
\begin{equation}
 \label{eq:masscons}
\frac{\partial \Sigma}{\partial t}+\frac{1}{r}\frac{\partial}{\partial r}(r v \Sigma)=0,
\end{equation}
and angular momentum
\begin{equation}
\begin{aligned}
\label{eq:angcons}
0 =&  \frac{\partial }{\partial t}\left(\Sigma r^2\Omega \boldsymbol{\ell}\right)
  +\frac{1}{r}\frac{\partial}{\partial r}(\Sigma v_r r^3\Omega \boldsymbol{\ell}) \\
  &-\frac{1}{r}\frac{\partial}{\partial r}\left(Q_1\Sigma c_{s}^2 r^2\boldsymbol{\ell}\right)
  - \frac{1}{r}\frac{\partial}{\partial r}\left(Q_2\Sigma c_{s}^2 r^3\frac{\partial \boldsymbol{\ell}}{\partial r}\right) \\
  &- \frac{1}{r}\frac{\partial}{\partial r}\left(Q_3\Sigma c_{s}^2 r^3\boldsymbol{\ell}\times\frac{\partial \boldsymbol{\ell}}{\partial r}\right),
  \end{aligned}
\end{equation}
respectively. The relevant viscosity coefficients have been rebranded as $Q_i$,  related to $\nu$ and $\nu_{\perp}$ through 
$Q_{1} = {\nu \Omega^\p }/{c_s^2}$ 
and $Q_{2} = {\nu_{\perp} \Omega}/{2 c_s^2}$ \cite{oglat13}. Using the warp-free, thin-disc structure relations \eqref{eq:diskrelns} 
within the definitions of $Q_{i}$ yields the leading-order terms: calculating the higher-order corrections requires an analysis of the (GR) 
Navier-Stokes equations, beyond the scope of this article (see Ref.~\cite{og99} for a Newtonian analysis).
It is easy, however, to show that the leading-order expressions for the two dissipative coefficients read (see equations 141--143 therein)
\begin{equation} \label{eq:q1}
Q_{1} \approx -\frac{3 \alpha}{2} + \frac{\alpha_{\perp}}{8} |\psi_0|^2 + \mathcal{O}(|\psi|^4),
\end{equation}
and
\begin{equation} \label{eq:q2}
Q_{2} \approx \frac{\alpha_{\perp}}{2} + \mathcal{O}(\psi^2),
\end{equation}
applying to the case of a polytropic disk with adiabatic index $\Gamma = 5/3$. For the third coefficient, we adopt the result from \citet{og99},
\begin{equation} \label{eq:q3}
Q_{3} \approx 3/8 + \mathcal{O}(|\psi|^2).
\end{equation}

The stability problem now boils down to considering perturbations of the fluid (e.g., $\Sigma$) and warp ($\psi_{0}$) variables to determine whether 
the eigenvalues of the perturbed variables indicate  growth or damping. Skipping details which can be found in \citet{dogan18}, linearisation 
of~\eqref{eq:angcons} eventually leads to the `coefficients determinant' equation
\begin{equation}
\label{eq:det}
\begin{vmatrix}
    s + 3 \alpha - \tfrac{1}{4}\psi_0^2 \alpha_{\perp}  & -\tfrac{1}{2} |\psi_0| \alpha_{\perp} & 0  \\
    5 \alpha \psi_0 - \tfrac{5}{12} \left( \psi_0^2 - 4 \right) \alpha_{\perp} & s +\alpha_{\perp}\left(1 - \tfrac{1}{2}\psi_0^2 \right) & - \tfrac{3}{8} \\
    \tfrac{5}{8}|\psi_0|   & \tfrac{3}{8} & s+  \alpha_{\perp}
\end{vmatrix} = 0,
\end{equation}
where we have made use of Eqs.~\eqref{eq:q1}--\eqref{eq:q3} (compare their equation 25).
The eigenvalue $s=-{i \omega}/{\Omega c_s^2 k^2}$ is defined such that a positive real part of $s$ indicates an unstable mode for wavenumber $k$. 
Eq.~\eqref{eq:det} reduces to a cubic in $s$, which can be solved easily given some relationship between $\alpha_{\perp}$ and $\alpha$, yielding three 
solutions $s_{i}(\alpha,\psi)$. One may then check the maximum of the real parts of $s_{i}$ to identify (in)stability.

\begin{figure}
\centering
 \includegraphics[width=0.48\textwidth]{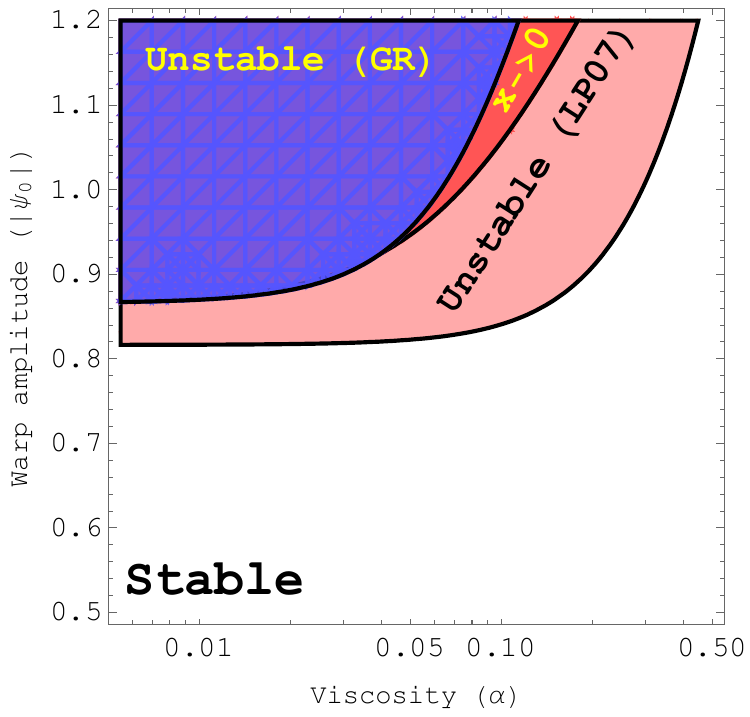}
 \caption{Stability diagram, as a function of viscosity and warp amplitude, for a highly-relativistic case with $x=1/6$ and $q=0.2$ (blue) and the Newtonian case ($x \to 0$; red). 
 For comparison, we also show the prediction obtained when using $\alpha_{\perp} = 1/\alpha$ (pink) from \protect\cite{lp07}.
 Shaded regions correspond to unstable cases with $\text{Re}(s) > 0$ from Eq.~\eqref{eq:det}.}
 \label{fig:instabhydro}
\end{figure}

Figure~\ref{fig:instabhydro} shows the instability region as a function of viscosity ($\alpha$) and warp ($|\psi_0|$) for a compact case with $x = 1/6$ and $q=0.2$ 
as compared to the Newtonian limit ($x \to 0$). Even for these relatively extreme parameter choices, we see that the two regions largely overlap except at large warps 
where the scheme breaks down. In general though, GR terms make the system more unstable: for $\psi_0 \approx 1$, a viscosity of $\alpha \lesssim 0.06$ 
leads to instability in the GR case, while $\alpha \lesssim 0.09$ is needed for a Newtonian system. 
This occurs because the ratio $\alpha_{\perp}/\alpha$ decreases when $r_0$ is small; see Fig.~\ref{fig:hydro_viscosities}. 
For comparison, we also show the predicted instability window using the LP07 relation $\alpha_{\perp} = 1/\alpha$; the window is wider in this case. 
This is somewhat unintuitive since the value of $\alpha_{\perp}$ is typically greater in this case for small $\alpha$, though occurs due to the opposing signs 
in the two terms within $Q_{1}$ \eqref{eq:q1}.  One can compare Fig.~\ref{fig:instabhydro} directly with Figure 5 from \citet{dogan18}, noting however that a 
numerical routine is used there to compute the $Q_{i}$ as functions of $\alpha$ and $\psi$ more accurately using the warped generalisation of the thin disc 
structure equations from \citet{og99} and subsequent works.

\section{Inclusion of a star-hosted magnetic field}
\label{sec:magfield}

In this section we reexamine various elements of the hydrodynamical case by including a magnetic field sourced by the central compact body.
The resulting model will be one of a magnetoviscous disc.    
Where convenient, we adopt the notation $B_{x} = B/(10^x\,\mbox{G})$ for the normalised magnetic field strength $B$.

\subsection{Torque comparison}
\label{sec:Btorque}

In the presence of a central body-hosted magnetic field, we expect the LT precession of adjacent fluid rings to be counteracted by a magnetic tension 
force due to the induced deformation in the field lines threading the disc. In order to scope out the parameter space where this effect could be important we 
repeat the back of the envelope calculation of Section~\ref{sec:tearing} by including the magnetic torque. 

The total magnetic torque is given by 
\be
N_{\rm B} \sim r^3 B_\varphi B_z,
\label{NB}
\ee
which assumes an equatorial disc threaded by a vertical magnetic field that is azimuthally dragged by the rotational  flow. 
We have omitted a constant geometric prefactor so that we can use the same approximate expression for the component of  $\mathbf{N}_{\rm B}$
along a given axis. For example,  given the relative space orientation of the stellar magnetic and the disc, we expect the torque component perpendicular 
to the disc to be comparable in magnitude to the equatorial component (cf. Ref.~\cite{gs21} and discussion therein).

For the precision of the present calculation it is sufficient to set $ B_\varphi \sim B_z$ while the the poloidal field is assumed to be a dipole 
$ B_z = -B_0 ( R/r)^3 $ where $B_0$ is the
field strength at the stellar surface (see Ref.~\cite{gs21} for a discussion). For the magnetic torque per unit mass, we find (again using the disk relations \ref{eq:diskrelns}) 
\be 
\label{eq:magtorque}
T_{\rm B}   \sim \frac{B_0^2} {4\pi \rho} \frac{R^3}{r^3}   \left ( \frac{H}{r} \right )^{-1}.
\ee
Inserting the density profile~\eqref{eq:density1} of a thin Keplerian disc, we can rewrite this as
\begin{align}
T_{\rm B}  &  \sim \frac{3}{4} \frac{\alpha B_0^2}{\dot{M}}   \left ( \frac{H}{r} \right )^2 \frac{ (GM)^{1/2} R^3}{r^{3/2}}
\\ 
& = \frac{9}{4} v_{\rm A}^2  \frac{c^3}{ G \dot{M}}   \left ( \frac{H}{r} \right )^2 \frac{\alpha}{\bar{r}^{3/2}},
\end{align}
where  $v_{\rm A}^2 = B_0^2 R^3 /3 M$ is the volume-averaged stellar Alfv\'en speed.
The magnetic field's backreaction would be unable to prevent the LT precession of the disc and the ensuing tearing when 
\be
T_{\rm LT} \gtrsim  T_{\rm B}.
\label{TorqueswithB1}
\ee
This leads to the following `magnetic tearing radius' condition 
\be
\bar{r}_{\rm t} \lesssim  \frac{q \sin\beta}{\alpha v^2_{\rm A}} \frac{G \dot{M}}{c} \left ( \frac{H}{r} \right )^{-2}.
\label{rteq2}
\ee
Therefore, the magnetic field threading the disc suppresses the onset of LT precession above some maximum distance $r_{\rm t}$ much alike
the viscous forces, see Eq.~\eqref{eq:tearing}.

We can obtain a more realistic result for the tearing radius by comparing the LT torque to the combined action of viscous and magnetic torques, i.e.
\be
T_{\rm LT} \gtrsim T_\perp +  T_{\rm B}.
\label{TorqueswithB2}
\ee
Eq.~\eqref{rteq2} is now replaced by 
\be
2 q \sin\beta  \left ( \frac{H}{r} \right )^{-2} \gtrsim \frac{9}{4}   \frac{c v_{\rm A}^2}{ G \dot{M}} \alpha\, \bar{r}^{1/2}+   | \psi | \alpha_\perp \bar{r}^{3/2},
\label{rteq3}
\ee
with $ | \psi| = \psi_0 $.
This is a cubic equation in $ \bar{r}^{1/2}$ and can be solved analytically; the resulting roots, however, are unwieldy. It is far more informative 
to plot~\eqref{rteq3} (as an equation) in the $B_0$-$\alpha$ plane for given values of the relevant stellar and accretion-specific parameters. 
For consistency however, this requires knowledge of how the $\alpha_{\perp}(\alpha)$ relation is modified by the magnetic field. Such a derivation 
is provided in Sec.~\ref{sec:magalphaperp}, where we explore the inequality \eqref{rteq3}. In any case, the two terms on the right-hand side of 
expression \eqref{rteq3} have a relative magnitude of
\begin{equation} 
\label{eq:ratiort}
\begin{aligned}
\left | \frac{c v_{\rm A}^2  \alpha \bar{r}^{1/2} / G \dot{M}}{ \psi  \alpha_\perp \bar{r}^{3/2}} \right |
\approx \,&  10^{3} \left(\frac{B_{0}}{10^{8} \text{ G}}\right)^{2} \left( \frac{\alpha}{\alpha_{\perp}} \right) \left(\frac{R}{r_{0}}\right) \\
&\times  \left(\frac{\dot{M}}{10^{-3} \dot{M}_{\rm edd}}\right)^{-1} \left(\frac{0.3}{|\psi|}\right).
\end{aligned}
\end{equation}
This shows that magnetic-adjustments to $\alpha_{\perp}$ itself are unlikely to play a significant role in tearing radii estimates, as the first term within~\eqref{rteq3} 
dominates except in systems with weak fields and large warps at large radii $r  \gg R$. For weak fields, however, the non-magnetic estimate \eqref{perpviscnewt} 
for $\alpha_{\perp}$  may be applied reliably. Note that if the disc is truncated beyond the geodesic innermost stable orbit thanks to the magnetic field (i.e., at the Alfv{\'e}n radius $R_{\rm A}$), 
one may expect that LT precession and disc-fragmentation would occur within the annulus $ R_{\rm A} \lesssim r \lesssim r_{\rm t}$. For even larger fields, the magnetosphere 
may be completely truncated (see Section 7 in Ref.~\cite{pat17}).

\subsection{Local analysis}
\label{sec:maglocal}

Our next task is to include the magnetic field dynamics in the fluid parcel test-body model of the previous sections to get $\alpha_{\perp}(\alpha,B)$. The only modification 
required is the addition in the equation of motion of the Lorentz force density,
\be
\bfL = \frac{1}{4\pi} \left [\, ( \bnabla \times \bB ) \times \bB \, \right ].
\ee
To do so, we orient some local Cartesian coordinates so that $\xhat = \hat{\mathbf{r}}$ and $\yhat = \hat{\boldsymbol{\varphi}}$ (the azimuth)
and assume a background magnetic field of the form
\be
\bB_0 (x)  = \left(\,0, B_y (x), B_z (x) \, \right),
\ee
which is trivially solenoidal. The motion of a fluid element off a circular equatorial orbit leads to the perturbed field 
\be
\mathbf{b} (t,x,z) = \bB - \bB_0 = (\, b_x, b_y, b_z  \, ).
\ee
For the background and perturbed Lorentz force we find, respectively,
\be
(\bfL)_0 = - \frac{1}{8\pi} \partial_x ( \bB_0^2 ) \xhat =  - \frac{1}{4\pi} \left ( B_y B_y^\p + B_z B_z^\p \right )  \xhat,
\ee
and 
\begin{align}
\delta \bfL &= \frac{1}{4\pi} \Big \{   \left [ B_z (\partial_z b_x - \partial_x b_z) - (b_z B^\p_z + b_y B^\p_y) -B_y \partial_x b_y \right ] \xhat 
\nonumber
\\ 
& +  (b_x B_y^\p + B_z \partial_z b_y) \yhat +  ( b_x B_z^\p- B_y \partial_z b_y) \zhat \Big \},
\end{align}
where a prime denotes a $d/dx$ derivative in this case.

The $\boldsymbol{B}$-field-augmented equations of motion are 
\begin{align}
& \delta \ddot{r}  + \Omega_r^2   \delta r + \lambda  \delta \dot{r}  =  H \Omega^2_0 \psi  
\nonumber \\
& \qquad +  \frac{1}{4\pi \rho}  \left [ B_z (\partial_z b_x - \partial_x b_z) - (b_z B^\p_z + b_y B^\p_y) -B_y \partial_x b_y \right ],
\label{equrad2}
\\  
&  \ddot{\psi}  + \Omega_\theta^2 \psi  + \lambda_\perp  \dot{\psi} =  \frac{1}{4\pi \rho r_0}   ( b_x B_z^\p- B_y \partial_z b_y).
\label{equz2}
 \end{align}
 The evolution of the magnetic field itself is described by the induction equation,
\be
\partial_t \bB = \bnabla \times ( \bv \times \bB ).
\ee
This is trivially solved by the background velocity, $\bv_0 = v_0 (x) \yhat$, and $\bB_0$ field; the perturbed equation is found to be
\begin{align}
\partial_t \mathbf{b} &= ( B_z \partial_z \delta v_x) \xhat  - \partial_x ( B_z \delta v_x ) \zhat
\nonumber 
\\ 
& +  \left [ \partial_x ( v_0 b_x - B_y \delta v_x ) + v_0 \partial_z b_z - B_y \partial_z \delta v_z \right ] \yhat,
\end{align}
where $\delta \bv = (\delta v_x, 0, \delta v_z )$. Expanding its components and using $ \bnabla \cdot \mathbf{b} = 0 $,
\begin{align}
\partial_t b_x & = B_z \partial_z \delta v_x,
\\ 
\partial_t b_z & = - B_z^\p \delta v_x  - B_z  \partial_x  \delta v_x,
\\ \nonumber  
\partial_t b_y & = v_0^\p b_x - B_y^\p \delta v_x   - B_y  ( \partial_x \delta v_x  + \partial_z \delta v_z ).
\end{align}
The first two expressions can be integrated in time and provides us with solutions for $b_x, b_z$ which can be subsequently inserted 
in the previous equations of motion:
\begin{align}
b_x = B_z \partial_z \delta r, \qquad  b_z = - \left ( B_z^\p \delta r  + B_z  \partial_x  \delta r \right).
\end{align}
The same procedure is not possible for $b_y$ unless we are prepared to increase the differential order of the equations of motion. 
In order to avoid the overcomplication of what is supposed to be a toy model, we remove all $b_y$ terms by setting $B_y =0$ (i.e.
we assume a \emph{purely vertical} background magnetic field; see Ref.~\cite{cel25} for a discussion).

The resulting equations of motion are
\begin{align}
& \delta \ddot{r}  + \left [  \Omega_r^2 - \frac{B_z B_z^{\pp}}{4\pi \rho} - \frac{ (B_z^\p)^2}{4\pi \rho}  \right ]   \delta r + \lambda  \delta \dot{r}  
=  H \Omega^2_0 \psi  
\nonumber \\
& \qquad +  \frac{1}{4\pi \rho}  \left [ B_z^2 \left ( \partial_z^2 \delta r + \partial^2_x \delta r \right ) + 3 B_z^\p B_z \partial_x \delta r \right ],
\label{equrad3}
\\  
&  \ddot{\psi}  + \Omega_\theta^2 \psi  + \lambda_\perp  \dot{\psi} =  \frac{B_z B_z^\p}{4\pi \rho r_0}  \partial_z \delta r.
\label{equz3}
 \end{align}
 {The local test-body model of the present analysis means that we should collapse the derivatives as $\partial_z \approx \pm 1/H,  \partial_x = \pm \partial_r \approx \pm \psi_0 / H$.
 This step unavoidably involves some loss of information related to the sign of each derivative (which in turn reflects an uncertainty in the directionality of the 
 relevant vectors in the problem, e.g., how the magnetic field is oriented with respect to the tilt). 
 In order to keep track of these signs we introduce the bookkeeping parameters  $\epsilon_i = \pm 1$ with $i=1,2, \ldots$. }
 
\begin{figure*}
\centering
 \includegraphics[width=0.7\textwidth]{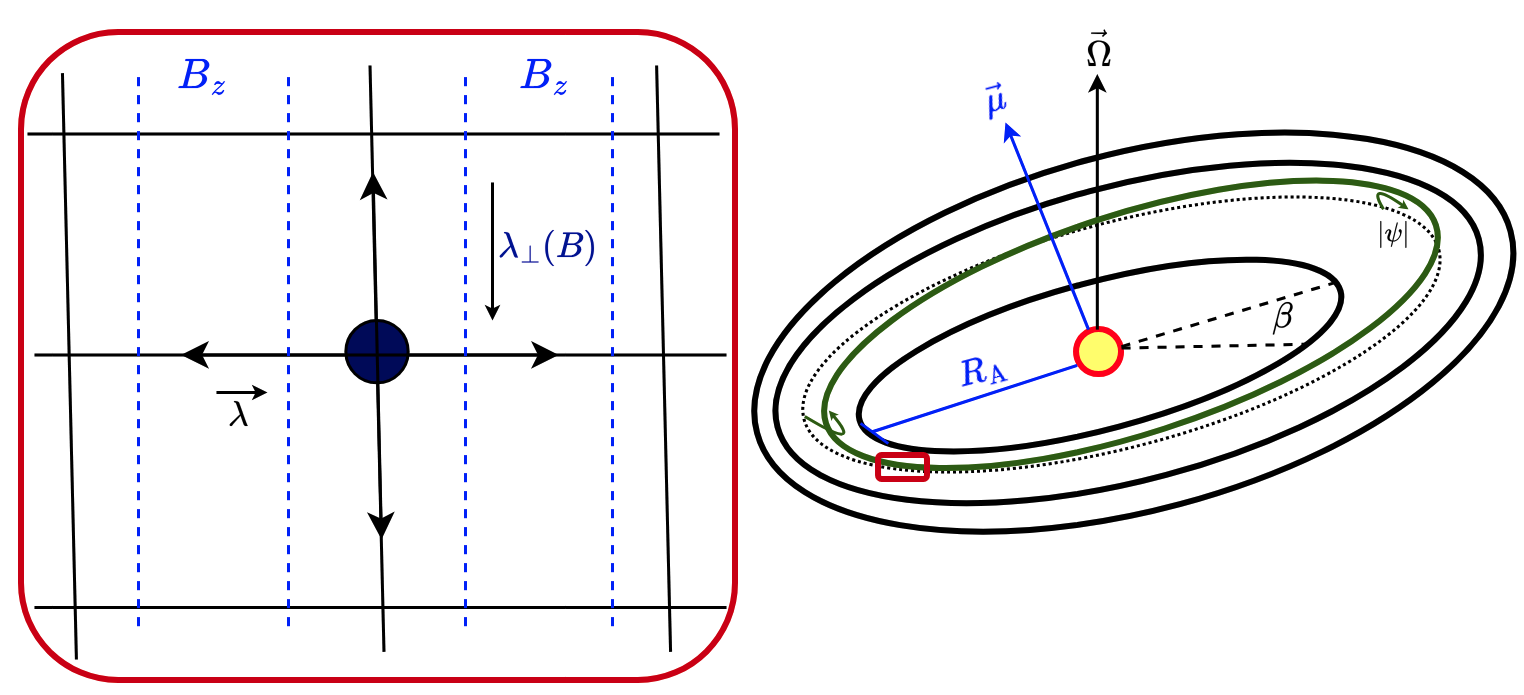}
 \caption{{Similar to Fig.~\ref{fig:schematic} though depicting the impact of a star-hosted magnetic field.
 If sufficiently magnetised, truncation of the disc occurs at the Alfv{\'e}n radius, $R_{\rm A}$, 
 where magnetic pressure balances the ram pressure of circulating material.
 The perpendicular viscosity, $\lambda_{\perp}$ increases sharply as a function of field strength above a certain cutoff (in the so-called `MAD' regime), adjusting the local parcel dynamics (see text).}}
  \label{fig:schematicMag}
\end{figure*}

 {While this simplification is somewhat unsatisfactory in that it introduces additional parameters in an already complicated system, it is not too difficult to see that
 each derivative term should take \emph{both} positive and negative values across different regions of the warped disc as a result of the oscillatory character of 
 the radial displacement $\delta r$, see Figs.~\ref{fig:schematic} and \ref{fig:schematicMag}. This is a key point because, as we will see, only positive signs can give rise to 
 instabilities, suggesting that some configurations warrant further investigation in the future via full MHD simulations.
In what follows, we therefore focus on regions of the disc where the $\epsilon$ are positive and investigate instability in those regions.  
 }

%
 
 We thus have the following algebraic expressions for the perturbed magnetic field:
\begin{align} 
\label{eq:fieldshear}
b_x  &= \epsilon_1 \frac{ B_z}{H} \delta r, \qquad  b_z  = -  \left ( B_z^\p  +  \epsilon_2 B_z  \frac{\psi_0}{H} \right  ) \delta r.
\end{align}
The equations of motion now take their final form
 \begin{align}
& \delta \ddot{r} + \tilde{\Omega}_r^2  \delta r   +  \lambda  \delta \dot{r}  =  H \Omega^2_0 \psi,
\label{equrad4}
\\   \nonumber 
\\
& \ddot{\psi} + \Omega^2_\theta \psi  + \lambda_\perp  \dot{\psi} =   \epsilon_1 \frac{B_z B_z^\p }{4\pi \rho H r_0}    \delta r,
\label{equz4}
 \end{align}
 with
 \begin{align}
 \tilde{\Omega}_r^2 & = \Omega_r^2 - \frac{1}{4\pi \rho} \Bigg [ ( B_z^\p )^2  + B_z B_z^{\pp}  +  \epsilon_3 \left ( \frac{B_z}{H} \right )^2 
 \nonumber
 \\ 
 &  + \epsilon_4 \psi_0^2  \left ( \frac{B_z}{H} \right )^2   + 3 \epsilon_2 B_z B_z^\p \frac{\psi_0}{H} \Bigg ].
 \label{OmradB1}
\end{align}
The key difference of this system compared to the one of Sec.~\ref{sec:local} is that the magnetic field couples the
two oscillators. Note that, modulo the magnetic shift of the orbital frequency, the radial equation is the same as in the 
non-magnetic model. The key difference appears in the vertical oscillator \eqref{equz4} which now contains a driving force 
term due to the magnetic coupling with the radial motion. 
{A schematic for the magnetoviscous dynamics, with a star-hosted dipole field but a locally-vertical one in a given fluid cell, is depicted in Fig.~\ref{fig:schematicMag}.}

 Our perturbative scheme requires $| \mathbf{b} | \ll B_0 $ which means 
 \be
 | b_x / B_z| \ll 1, | b_z/B_z| \ll 1.
 \ee
This, combined with the derivatives approximation, places an upper limit to the radial displacement
\be
| \delta r | \ll \{r_0,  H, H/\psi_0 \} ~\Rightarrow ~ | \delta r | \ll H.
\ee
 We proceed to solve the system~\eqref{equrad4}-\eqref{equz4} by means of the normal-mode ansatz $\delta r = \cA_r e^{i \sigma t},~\psi = \psi_0 e^{i \sigma t}$, 
 with constant amplitudes $\cA_r$ and  $\psi_0$. The resulting algebraic system
 \begin{align}
& \left ( -\sigma^2  + \tilde{\Omega}_r^2  + i \sigma  \lambda \right ) \cA_r  =  H \Omega^2_0 \psi_0,
\label{equsys1}
\\  
& \left ( -\sigma^2 + \Omega^2_\theta   + i \sigma  \lambda_\perp \right ) \psi_0 =   \epsilon_1  \frac{B_z B_z^\p }{4\pi \rho H r_0} \cA_r,
\label{equsys2}
 \end{align}
 admits a non-trivial solution provided the determinant of the system's matrix vanishes, i.e.
 \be
  ( \sigma^2  - \tilde{\Omega}_r^2  - i \sigma  \lambda )  ( \sigma^2 - \Omega^2_\theta - i \sigma  \lambda_\perp )
  =   \epsilon_1 \frac{B_z B_z^\p }{4\pi \rho r_0}  \Omega^2_0 .
  \label{detomega}
 \ee
 In the non-magnetic limit this equation returns the viscosity-modified orbital frequencies (plus viscous damping) of Section~\ref{sec:local}:
 \begin{align}
 \sigma_r & =  i \lambda/2 \pm  \omega_r  = i \lambda/2 \pm ( \Omega_r^2 -  \lambda^2/4)^{1/2}, 
 \label{noBom1}
 \\ 
 \sigma_\theta &=  i \lambda_\perp/2 \pm  \omega_\theta = i \lambda_\perp/2 \pm \left (  \Omega_\theta^2 - \lambda_\perp^2/4 \right )^{1/2}. 
 \label{noBom2}
 \end{align}
 Recall that, in that earlier section, we only considered the $\sigma_\theta$ mode as the one driven by warp. In the present model,
 the coupling between the oscillators renders both modes relevant. 
 
 More interesting for the discussion of this section are the \emph{magnetic field-dominated} modes. These modes, which we denote as $\sigma_{\rm B}$, 
 should appear when the viscosity terms in~\eqref{detomega} are negligible compared to the magnetic terms. After setting $\lambda, \lambda_\perp \to 0$
 we find the roots
 \be
 \sigma_{\rm B}^2 = \frac{1}{2} \left [ \, \tilde{\Omega}^2_r + \Omega_\theta^2 \pm 
 \sqrt{\left ( \tilde{\Omega}^2_r - \Omega_\theta^2 \right )^2 + \epsilon_1 \frac{B_z B_z^\p}{\pi \rho r_0} \Omega_0^2} \, \right ].
   \label{Bom1}
 \ee
 This result can be further expanded if we first  define the disc's  \emph{local} Alfv\'en speed 
 defined as
 \be
 c_{\rm A}^2 = \frac{B_z^2}{4\pi \rho},
 \ee
and subsequently rewrite~\eqref{OmradB1} as
 \begin{align}
 \tilde{\Omega}_r^2 & = \Omega_r^2 - \Omega_0^2 \gamma  \Big [ \left  ( \frac{r_0  B_z^\p}{B_z} \right )^2  +  \frac{r_0^2 B_z^{\pp}}{B_z}  
 +  \epsilon_3 \left ( \frac{r_0}{H} \right )^2 
 \nonumber
 \\ 
 &  + \epsilon_4 \psi_0^2  \left ( \frac{r_0}{H} \right )^2   + 3 \epsilon_2 \psi_0 \frac{r_0  B_z^\p}{B_z} \frac{r_0}{H} \Big ]
 \equiv \Omega^2_r - \Omega_0^2 \tilde{\omega}_{\rm B},
 \label{OmradB2}
\end{align}
where we have defined the dimensionless parameters $ \tilde{\omega}_{\rm B}$ and 
\be
\gamma = \left ( \frac{c_{\rm A}}{r_0\Omega} \right )^2. 
\ee
With the help of the additional approximations from \eqref{eq:omegaexp}
\be
 \Omega_r^2 - \Omega_\theta^2 \approx 6 x \Omega_0^2, \quad \Omega_\theta \approx \Omega_0,
 \label{Omapprox1}
 \ee
 we find
 \be
 \left ( \frac{\sigma_{\rm B}}{\Omega_0} \right )^2 =   1 + 3 x - \frac{1}{2}  \tilde{\omega}_{\rm B} \pm  
 \sqrt{  \frac{1}{4}  \tilde{\omega}_{\rm B}^2 -  3 x  \tilde{\omega}_{\rm B}  +    \epsilon_1 \gamma \frac{r_0 B_z^\p}{B_z} }. 
  \label{Bom2}
 \ee
 The dominant term under the square root is the first one whenever $ |\tilde{\omega}_{\rm B} | \gg 1$. Given that  $ | r_0 B_z^\p / B_z | \sim 1$
 this condition translates to $\gamma (r_0 /H )^2 \gg 1$.
 After expanding the square root and dropping the relativistic terms we find
 \be
 \left ( \frac{\sigma_{\rm B}}{\Omega_0} \right )^2 \approx  1  - \frac{1}{2}   \tilde{\omega}_{\rm B}   \pm  \frac{1}{2}  \tilde{\omega}_{\rm B} 
  \pm    \frac{\epsilon_1 \gamma}{ \tilde{\omega}_{\rm B}} \frac{r_0 B_z^\p}{B_z},
  \label{Bom3}
 \ee
 where the sign choice in the last two terms is independent. Of particular interest is the (lower sign) mode
 \be
  \sigma_{\rm B}^2 \approx - \Omega_0^2 \tilde{\omega}_{\rm B} \approx  -\epsilon_3 \Omega_0^2 \gamma  \left ( \frac{r_0}{H} \right )^2
  =  -\epsilon_3 \left ( \frac{c_{\rm A}}{H} \right )^2,
 \label{Bom4}
 \ee
 which could signal the onset of an instability {in regions of the disc where} $\epsilon_3>0$. In contrast, the upper sign choice leads to `hydrodynamical modes'
 \be
 \left ( \frac{\sigma_{\rm B}}{\Omega_0} \right )^2 \approx  1 \pm    \frac{\epsilon_3 \gamma}{ \tilde{\omega}_{\rm B}} \frac{r_0 B_z^\p}{B_z},
  \label{Bom5}
 \ee
 which are slightly modified by the magnetic field. Therefore, only expression~\eqref{Bom4} represents the magnetic-field dominated modes. 
 
 With the result~\eqref{Bom4} at hand,  we can loop back to our previous discussion (see paragraph before Eq.~\ref{Bom1}) 
 and quantify the necessary condition for the omission of the viscous terms. This is
  \be
  | \sigma_{\rm B} |  \gg ( \lambda, \lambda_\perp ) ~\Rightarrow ~  \frac{c_{\rm A}}{H} = \gamma^{1/2} \frac{r_0}{H} \gg ( \alpha, \alpha_\perp).
 \label{eq:gammaeqn0}
 \ee
Combined with the condition $ | \tilde{\omega}_{\rm B}| \gg 1$ used earlier (and the standard assumption $\alpha \ll 1$) we can write the
following `global' condition for the emergence of the magnetic modes:
\be
\gamma \left ( \frac{r_0}{H} \right )^2 \gg ( 1, \alpha_\perp^2).
\label{eq:gammaeqn1}
\ee 
 Note that $\gamma$ is also a proxy for the ratio of magnetic to fluid energy density (or pressure) of the unperturbed disc;  we must demand 
 $c_{\rm A} < v_\varphi$, otherwise the magnetic pressure would be strong enough as to cause the disruption of the disc. In particular, $\gamma = 1$ marks
 the location of the Alfv{\'e}n radius\footnote{As we have seen, the presence of a warp $\psi$ perturbs the magnetic field as 
 $B_{z} \to B_{z} \left [ 1 + 2 (\psi /H ) \cA \sin \omega_{\theta} t \right ]$ 
 which means that $ R_{\rm A} $ should adjust accordingly. Such an oscillating field can lead to an effect known as `magnetic pumping', accelerating particles 
 reaching the Alfv{\'e}n surface in a direction that depends on the phase of the oscillation \cite{kiuj97}. Transient phenomena at frequencies of $\sim \omega_{\theta}$ 
 with amplitudes depending on $\psi/H$ would thus be expected, which could explain the $\sim$~kHz quasi-periodic oscillations seen in a number of LMXBs \cite{kils98}.}
 \be
 \label{eq:alfven}
 R_{\rm A} = \xi \frac{ B_0^{4/7} R^{12/7}}{\dot{M}^{2/7} (G M)^{1/7}} ,
 \ee
with\footnote{There is an inconsequential typo in expression (5) from Ref.~\cite{gs21}, where the \citet{ss73} thin-disc $\xi$ is reported with 
a slightly larger numerical prefactor. Note, however, that $\xi$ is often treated as a free parameter.} 
$\xi \approx \left( 3 \alpha/4 \right)^{2/7} \left({H}/{R_{\rm A}}\right)^{6/7}$ from Eq.~\eqref{eq:density1}. Based on this discussion we can easily obtain, for example, 
the radius at which $\gamma (r_0) ( r_0/ H)^n = 1$. This is equivalent to the Alfv\'en radius expression with a rescaled magnetic field $B_0 \to B_0 (r_0/H)^{n/2}$, leading
to $r_0 = R_{\rm A}  ( R_{\rm A}/H )^{2n/7} $. Therefore, the condition~\eqref{eq:gammaeqn1} is satisfied within the annular region 
\be \label{gammaalt}
R_{\rm A} \lesssim r_ 0 \ll R_{\rm A}  \left (  \frac{R_{\rm A}}{\alpha_\perp H } \right )^{4/7}.
\ee
This condition obviously implies an upper limit on the vertical viscosity, namely, $\alpha_\perp <  R_{\rm A} / H $. 

\begin{figure}
\centering
 \includegraphics[width=0.48\textwidth]{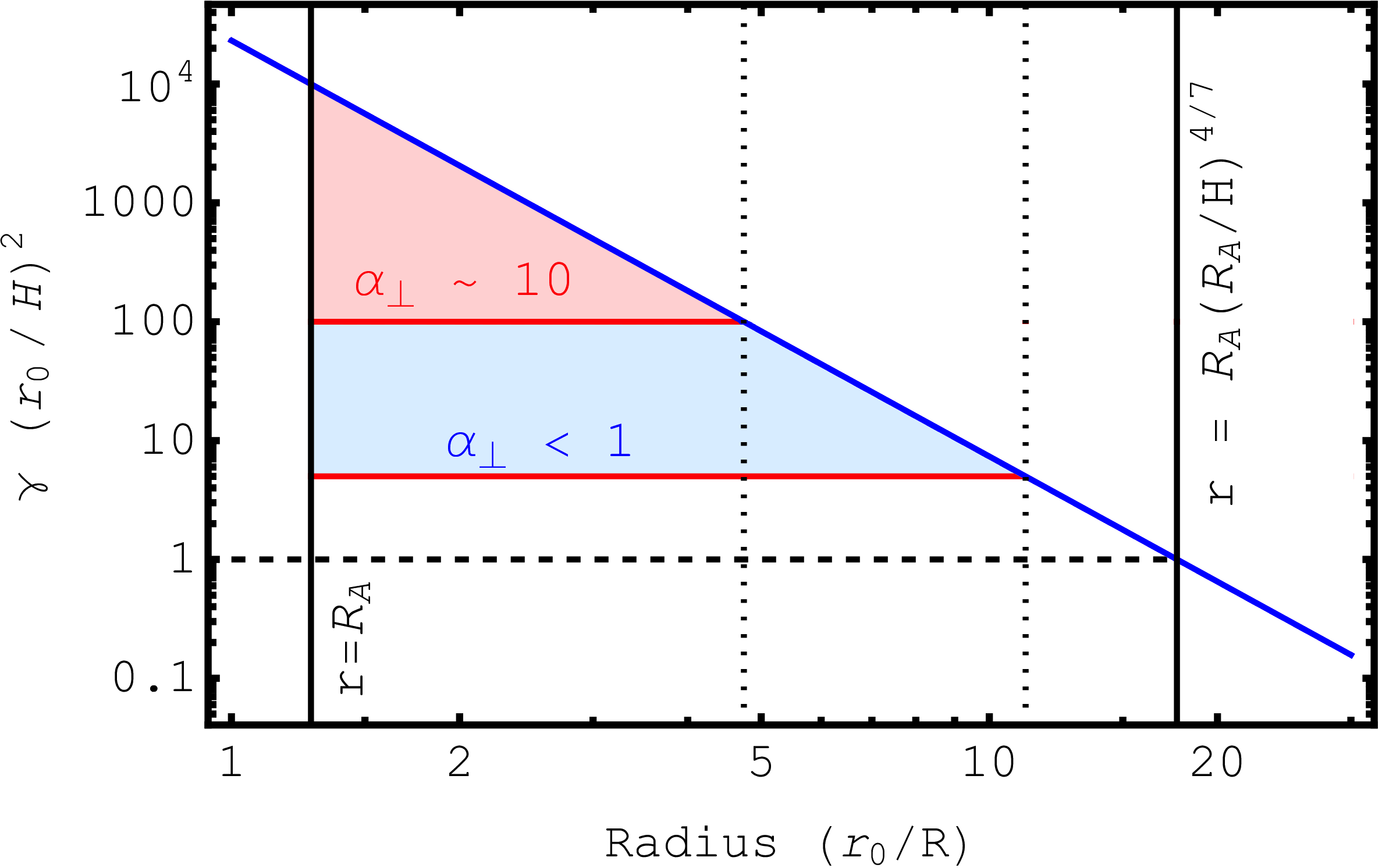}
 \caption{Dimensionless magnetic parameter $\gamma (r_{0}/H)^2$ from expression \eqref{gammaprofile} as a function of radius for $H/r_{0} = 10^{-2}$, 
 $\alpha = 0.1$, $R = 10$~km, $M = 1.4 M_{\odot}$, $B_{0} = 2 \times 10^{8}$~G, and $\dot{M} = 10^{-6} \dot{M}_{\rm edd}$ (solid blue curve). The solid, vertical lines 
 depict the limiting values associated with total disc truncation ($\gamma = 1$; left) and equality in \eqref{gammaalt} (right) for $\alpha_{\perp} = 1$. The dotted, vertical 
 lines mark the critical transition radius associated with the emergence of magnetic modes from condition \eqref{eq:gammaeqn1} for either $\alpha_{\perp} \sim 1$ 
 (with `$\gg$' interpreted as five in this case; blue shaded region) or $\alpha_{\perp} \sim 10$ (red shaded region). The dashed, horizontal line marks 
 $\gamma (r_{0}/H)^2 = 1$. }
 \label{fig:gammafn}
\end{figure}

The parameter $\gamma$ clearly plays an important role in the dynamics of a magnetoviscous disc. Using the 
thin disc density profile~\eqref{eq:density1} we can write it as ({for a dipole}),
\be
\gamma = \frac{3}{4} \frac{\alpha}{\dot{M}} \left ( \frac{H}{r_0} \right )^3 \frac{B_0^2 R^6}{r_0^5 \Omega_0}.
\label{gammaprofile}
\ee
The function $\gamma (r_0)$ is depicted in Fig.~\ref{fig:gammafn} for parameter choices appropriate for neutron star LMXBs 
(see also Sec.~\ref{sec:lmxbs}), together with a number of relevant radii and regions described in this section.

In the presence of both viscosity and magnetic field the full mode  equation~\eqref{detomega} must be solved {to get the eigenfrequencies $\sigma$}. 
Ignoring the relativistic corrections in the frequencies, $\Omega_r \approx \Omega_\theta \approx \Omega_0$, we can write~\eqref{detomega}
 in the following dimensionless form
\begin{align}
\hspace{-0.2cm} \left ( \tilde{\sigma}^2 -1 + \tilde{\omega}_{\rm B} -i \alpha \tilde{\sigma}   \right ) \left ( \tilde{\sigma}^2 - 1 -i \alpha_\perp \tilde{\sigma}  \right ) 
= \epsilon_1 \gamma \left ( \frac{ r_0 B^\p_z}{B_z}\right ),
\label{detomega2}
\end{align}
where $\tilde{\sigma} = \sigma/\Omega_0$.  This quartic equation can be solved analytically but the resulting expressions are too cumbersome. 
Nevertheless, the existence of unstable modes can be diagnosed by transforming $\tilde{\sigma} \to - i \tilde{\varsigma}$ so that~\eqref{detomega2} becomes a 
fourth-order polynomial in $\tilde{\varsigma}$ with real coefficients, from which the standard Routh-Hurwitz criteria apply for the existence of roots with negative real 
component~\cite{routh91}. Noting the positivity and smallness of the viscosities, we expect that the Hurwitz criteria can be violated only if the constant term in the 
polynomial is negative. That is, if
\be
\tilde{\omega}_{\rm B}  +  \epsilon_1 \gamma \left ( \frac{ r_0 B^\p_z}{B_z}\right ) > 1 
\ee
then there will exist an unstable branch.  To leading order this reduces to the simple condition
\be
\label{eq:notMRI}
\epsilon_1 \gamma  \left({r_0}/{H}\right)^{2}  > 1.
\ee
{This can only hold for positive $\epsilon_1$ in which case it is easy to see, using the structure equations \eqref{eq:diskrelns}, that it is 
equivalent to $c_{\rm A} > c_{s}$. This inequality marks the onset of a magnetically-dominated {(`MAD')} regime with plasma $\beta_{\rm plasma} < 1$. 
Again we emphasise that the signs of the $\epsilon_i$ parameters are expected to vary across the disc; in regions where the induced magnetic field is of the opposite polarity 
in Eq.~\eqref{eq:fieldshear} (i.e., $\epsilon_{1} < 0$), the system is instead stable as  perturbations are restorative.}


\begin{figure}
\centering
 \includegraphics[width=0.48\textwidth]{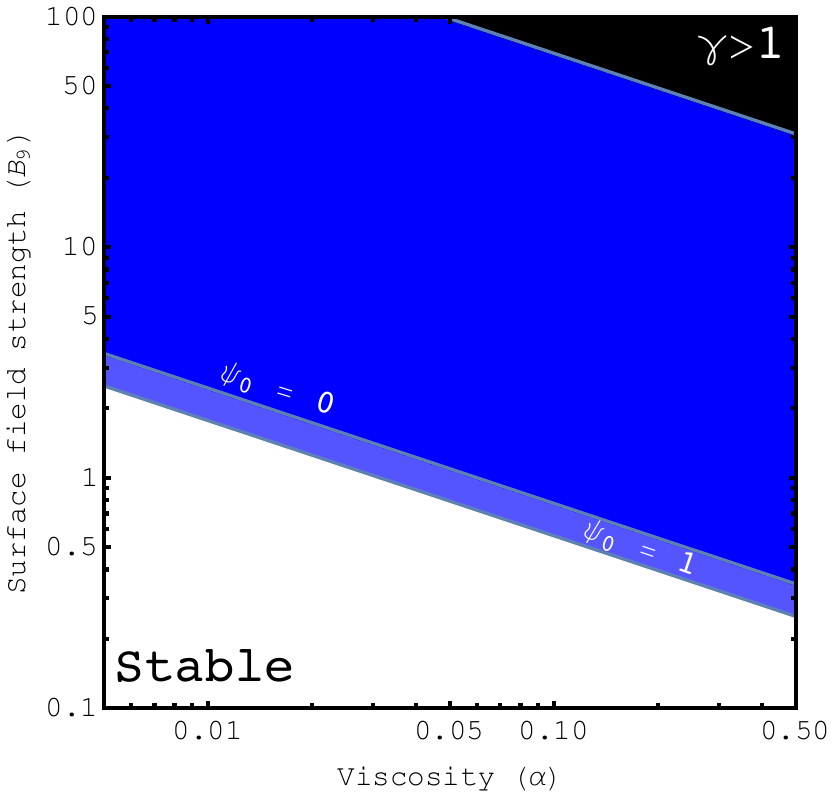}
 \caption{Stability diagram for $\alpha_{\perp} = 10 \alpha$, $H/r = 10^{-2}$, and $\dot{M} = 10^{-4} \dot{M}_{\rm Edd}$ at a  
 radius of $r = 5 R$ in the $B_{0}$-$\alpha$ plane for unwarped ($\psi_{0} = 0$; blue) or highly-warped ($\psi_{0} = 1$; purple) cases. We 
 fix $\epsilon_{i} = 1$ for all $i$. ({If negative, the entire domain remains stable as discussed in the main text.}) Shaded regions correspond to unstable cases with $\text{Im}(\sigma) < 0$. A visually-indistinguishable diagram is 
 obtained for any choice $0.1 \lesssim \alpha_{\perp} / \alpha \lesssim 10$. The region shaded in black corresponds to $\gamma > 1$, where the scheme 
 breaks down.}
 \label{fig:instab}
\end{figure}

The validity of condition \eqref{eq:notMRI} can be investigated numerically by simply computing the roots directly from Eq.~\eqref{detomega2} and checking 
if any satisfy $\text{Im}(\sigma) < 0$.  The result is shown in Fig.~\ref{fig:instab} in the $B_{0}$-$\alpha$ plane for some fiducial choices 
$\alpha_{\perp} = 10 \alpha$ and $\dot{M} = 10^{-4} M_{\rm Edd}$ at a radius of $r_0 = 5 R$ for either unwarped ($\psi_{0} = 0$) or highly-warped ($\psi_{0} = 1$) cases. 
We see that the stability boundary for the numerical solutions trace a straight line corresponding to constant values of $B_{0}^2 \alpha$. 
This follows from the combined proportionalities in $\rho \propto 1/\alpha$ and $c_{\rm A} \propto B_{0}^2/\rho$, indicating the validity of inequality~\eqref{eq:notMRI} 
as a diagnostic with small deviations provided by the warp parameter (at most a factor $\lesssim 2$ reduction to the critical $B_{0}$ for a given $\alpha$).
The fact that a magnetically-dominated transition corresponds precisely to the existence of unstable warp modes would appear to imply that episodic accretion 
is inevitable around stars with strong fields ($B_{0} \gtrsim 10^{9}$~G) unless the disc is very inviscid ($\alpha \ll 10^{-2}$), as theoretically expected 
for `magnetically-arrested disks' \cite{bis76,nar03}. Whether a more realistic magnetic geometry involving a toroidal field would adjust this conclusion is unclear; 
such an extension, together with astrophysical implications, will be investigated elsewhere.

\subsection{The $\alpha_\perp (\alpha)$ relation revisited} \label{sec:magalphaperp}

The next step of this analysis is the calculation of the velocities $\delta v_z \approx r_0 \dot{\psi},~\delta v_r = \delta \dot{r} $ and 
the use of the viscous dissipation balance~\eqref{dissbalance} in order to arrive to a revised relation $\alpha_\perp (\alpha)$ for a warped 
magnetoviscous disc. 
From Eq.~\eqref{equsys1} we have
\be
\cA = | \cA_r |  = \frac{H \Omega_0^2 \psi_0}{ \sqrt{ (\sigma_{\rm R}^2 -\sigma_{\rm I}^2 - \tilde{\Omega}^2_r + \sigma_{\rm I} \lambda )^2  
+ \sigma_{\rm R}^2 (2 \sigma_{\rm I} -\lambda)^2 }},
\label{AmplitB1}
\ee
where $\sigma_{\rm R}/\sigma_{\rm I}$ is a shorthand for the real/imaginary part of $\sigma$. For the non-magnetic modes~\eqref{noBom1}-\eqref{noBom2}
this expression returns the amplitude~\eqref{eq:hydroamp}. 

Using, as before, $ \partial_r \delta v_z \approx \dot{\psi}$ and $\partial_z \delta v_r \approx \delta \dot{r} /H$, \eqref{dissbalance} yields
\begin{align}
\frac{\alpha_\perp}{\alpha} & \approx \left ( \frac{\cA}{\psi_0 H} \right )^2 
\nonumber \\
&=  \frac{\Omega_0^4}{  (\sigma_{\rm R}^2 -\sigma_{\rm I}^2 - \tilde{\Omega}^2_r + \sigma_{\rm I} \alpha \Omega_0 )^2  + \sigma_{\rm R}^2 (2 \sigma_{\rm I} -\alpha \Omega_0)^2  }.
\label{a2awithB1}
\end{align}
This is an implicit expression for $\alpha_\perp (\alpha)$ because $\sigma$ itself is, in general, a function of the alpha parameters. Ignoring relativistic corrections
to the orbital frequencies, we can rewrite~\eqref{a2awithB1} as
\be
\frac{\alpha_\perp}{\alpha} \approx   \Big [  \Big ( \tilde{\sigma}_{\rm R}^2  -1 
+ \tilde{\omega}_{\rm B} + \tilde{\sigma}_{\rm I} ( \alpha -\tilde{\sigma}_{\rm I} )  \Big )^2 + \tilde{\sigma}_{\rm R}^2 (2 \tilde{\sigma}_{\rm I} - \alpha)^2 
\Big ]^{-1},
\label{a2awithB2}
\ee
where, as before, we have normalised the frequencies as $\tilde{\sigma}_{\rm R, I} = \sigma_{\rm R,I}/\Omega_0$. 
These expressions clearly suggest that in the strong-field regime of the disc (and especially in the vicinity of the Alfv\'en radius $R_{\rm A}$) the 
presence of the magnetic field will significantly modify the simple $\alpha_\perp (\alpha)$ relations of Section~\ref{sec:alphas}. 

\begin{figure}
\centering
 \includegraphics[width=0.48\textwidth]{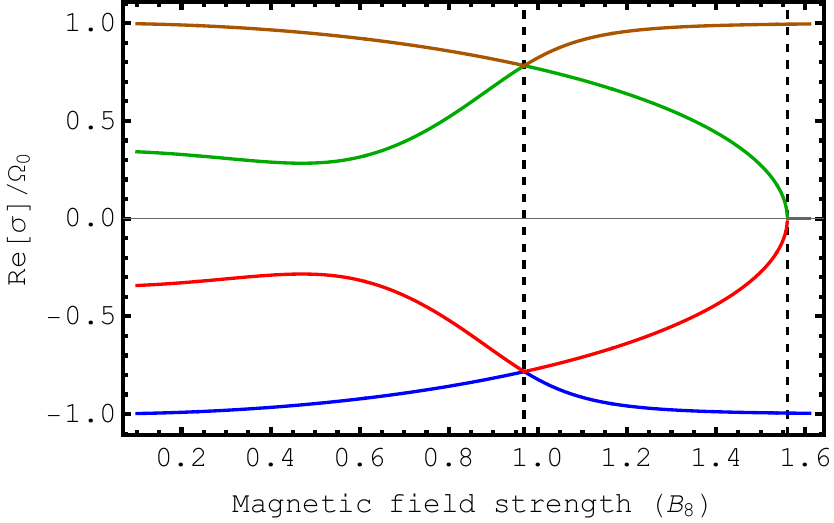}
 \caption{Real parts of normalised mode frequencies, $\tilde{\sigma}_{\rm R}$, determined by solving equations \eqref{detomega} and \eqref{a2awithB1} 
 simultaneously as a function of $B_{8} = B_{0}/10^{8}$~G, of which there are four in general (coloured curves). We fix $M = 1.4 M_{\odot}$, $R = 10$~km, 
 $\dot{M} = 10^{-4} \dot{M}_{\rm edd}$, $\psi_0 = 0.3$, $H/r_{0} = 10^{-2}$, $\alpha = 0.1$, and consider a radius $r_{0} = 2 R$ for demonstration. 
 The solid, vertical lines mark the onset  of avoiding crossings: tracing the colors (i.e., green to brown for $\tilde{\sigma}_{\rm R} >0$ or red to blue for 
 $\tilde{\sigma}_{\rm R} <0$) illustrates the changing character where the derivative of the eigenfrequency with respect to $B_{0}$ flips sign. 
 At high magnetic field strengths, two of the modes become imaginary and instead avoid a crossing in $\tilde{\sigma}_{\rm I}$. }
 \label{fig:avoided}
\end{figure}

The previously calculated inviscid magnetic modes $\sigma_{\rm B}$ are not suitable for a proper study of~\eqref{a2awithB2} as they cancel out to leading order
with the $\tilde{\omega}_{\rm B}$ term. The fact that the dispersion relation~\eqref{detomega} and expression~\eqref{a2awithB1} 
(or~\ref{detomega2} and~\ref{a2awithB2}) are coupled
necessitates a simultaneous solution to determine both the mode 
frequencies and magnetically-augmented $\alpha_{\perp}(\alpha)$ relation in the general case. The problem is non-trivial however, as the modes exhibit avoided-crossings; 
this is shown in Fig.~\ref{fig:avoided}, depicting the real component $\tilde{\sigma}_{\rm R}$, for some representative disc-star parameters. 
In this example, we see that already around $B_{0} \approx 10^{8}$~G the two branches residing in the upper-half plane meet and subsequently swap character, 
with the same occuring for those in the lower plane. After this point, the branch of relevance concerns that which asymptotes to the magnetically-dominated 
case~\eqref{Bom3}, so care must be taken as the crossing point varies as a function of radius for any given disc setup. Moreover, even before the transition point 
some mode frequencies become significantly affected by the magnetic pressure and increase in magnitude, highlighting the importance of magnetic 
corrections to the $\alpha_{\perp}/\alpha$ relation. For example, at $B_{0} \approx 10^{8}$~G we have that the lower-branches of $|\tilde{\sigma}_{\rm R}|$ increase 
in frequency by a factor $\approx 2$ relative to the non-magnetic limit. A second transition occurs at higher field strengths, where two of the modes become purely 
imaginary (around $B_{0} \approx 1.5 \times 10^{8}$~G in this example).

\begin{figure}
\centering
 \includegraphics[width=0.48\textwidth]{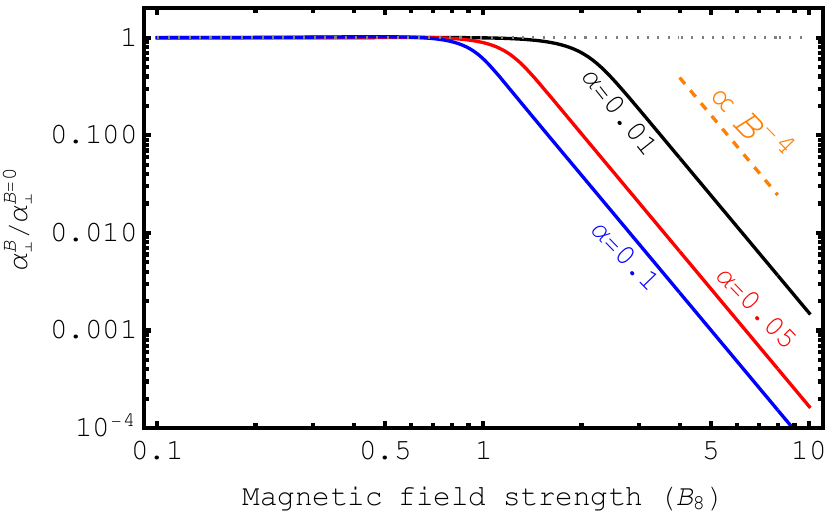}
 \includegraphics[width=0.48\textwidth]{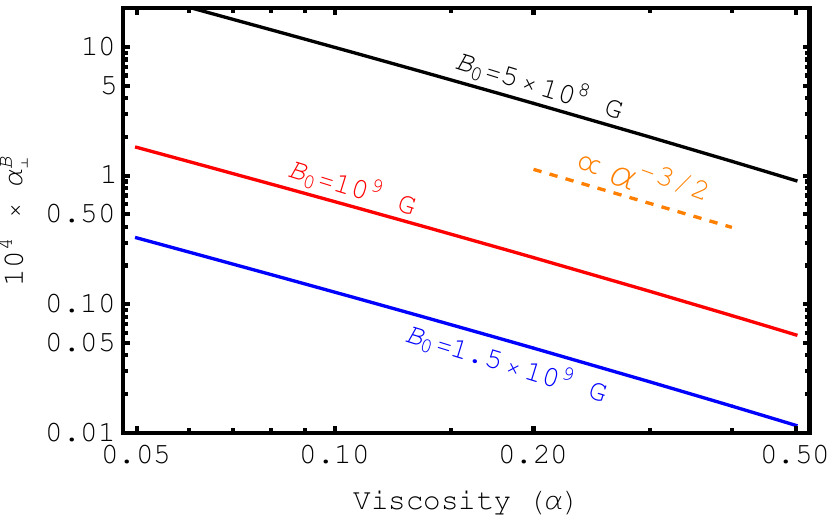}
 \caption{(Top panel): Ratio of the magnetically-adjusted perpendicular viscosity to the non-magnetic limit as a function of $B_{8}$ for a few representative 
 values of $\alpha$ (solid curves). The dotted line marks unity, demonstrating convergence for weak fields. The dashed line depicts a $B^{-4}$ falloff, which 
 all cases succumb to after when $B$ is large enough. (Bottom panel): Values of $\alpha_{\perp}^{\rm B}$ for fixed field strengths (solid curves) as a function 
 of $\alpha$. The dashed line in this case depicts a $\alpha^{-3/2}$ falloff; clearly all cases follow this behaviour also, noting the strong fields chosen such that 
 the magnetic branch resides within the entire parameter space shown. We set $H/r = 10^{-2}$, $\dot{M} = 10^{-4} \dot{M}_{\rm edd}$, 
 $R = 10^{6}$~cm, $M = 1.4 M_{\odot}$, at a fixed radius $r_{0} = 2 R$. 
 }
 \label{fig:perp_viscosity1}
\end{figure}

Accounting for this phenomenon via a numerical routine that selects the magnetically-dominated branch after the transition point, the adjusted $\alpha_{\perp}(\alpha)$ 
relation is shown in Fig.~\ref{fig:perp_viscosity1} as a function of $B_{0}$ (top panel) or $\alpha$ (bottom).

When $\alpha$ is fixed, we normalise the result by the non-magnetic limit to show the convergent behaviour; as expected, for $B_{0} \ll 10^{8}$~G  the ratio 
is numerically indistinguishable from unity. As the field strength increases and $\tilde{\Omega}^2_{r} - \Omega^2_{r}$ becomes large though we see a reduction 
in accord with expectations from the fact that large fields can make the system more unstable (Fig.~\ref{fig:instabhydro}). For example, at values of $\alpha \sim 10^{-2}$ 
we have $\alpha_{\perp}^{\rm B} / \alpha_{\perp}^{\rm B=0} \sim 10^{-3}$ for field  strengths $B_{0} \sim 10^{9}$~G -- a value not atypical for mature neutron stars 
in tight binaries (see, e.g., Ref.~\cite{gs21}). 

In the bottom panel, we instead plot the value of $\alpha_{\perp}$ directly without normalising by the non-magnetic limit. 
This is done to illustrate the clear fall-off behaviour: when the field is sufficiently strong, such that the magnetically-dominated branch applies, we find 
$\alpha_{\perp}^{\rm B} \propto 1/\alpha^{3/2}$. This indicates that strong magnetic fields, in addition to reducing the perpendicular viscosity like $B^{-4}$ (top panel), 
lead to a further, indirect reduction as $\alpha \ll 1$ and the scaling departs significantly from the non-magnetic \eqref{alphaperpN} and GR \eqref{alphaperpGR} cases.
For canonical neutron-star like parameters, tiny values of $\alpha^{\rm B}_{\perp} \lesssim 10^{-3}$ may be anticipated.

Taken together, the above numerical investigations suggest the following scaling in the MAD regime:
\be \label{eq:magviscreln}
\left (\frac{\alpha_\perp}{\alpha} \right )_{\rm B} \sim \frac{1}{\alpha^{5/2} \tilde{\omega}^2_{\rm B}} \sim  \frac{1}{\alpha^{5/2} B_0^4},
\ee
where the appearance of $\tilde{\omega}_{\rm B}$ is to be expected by its leading-order weight in the mode dispersion relation. 

\begin{figure}
\centering
 \includegraphics[width=0.48\textwidth]{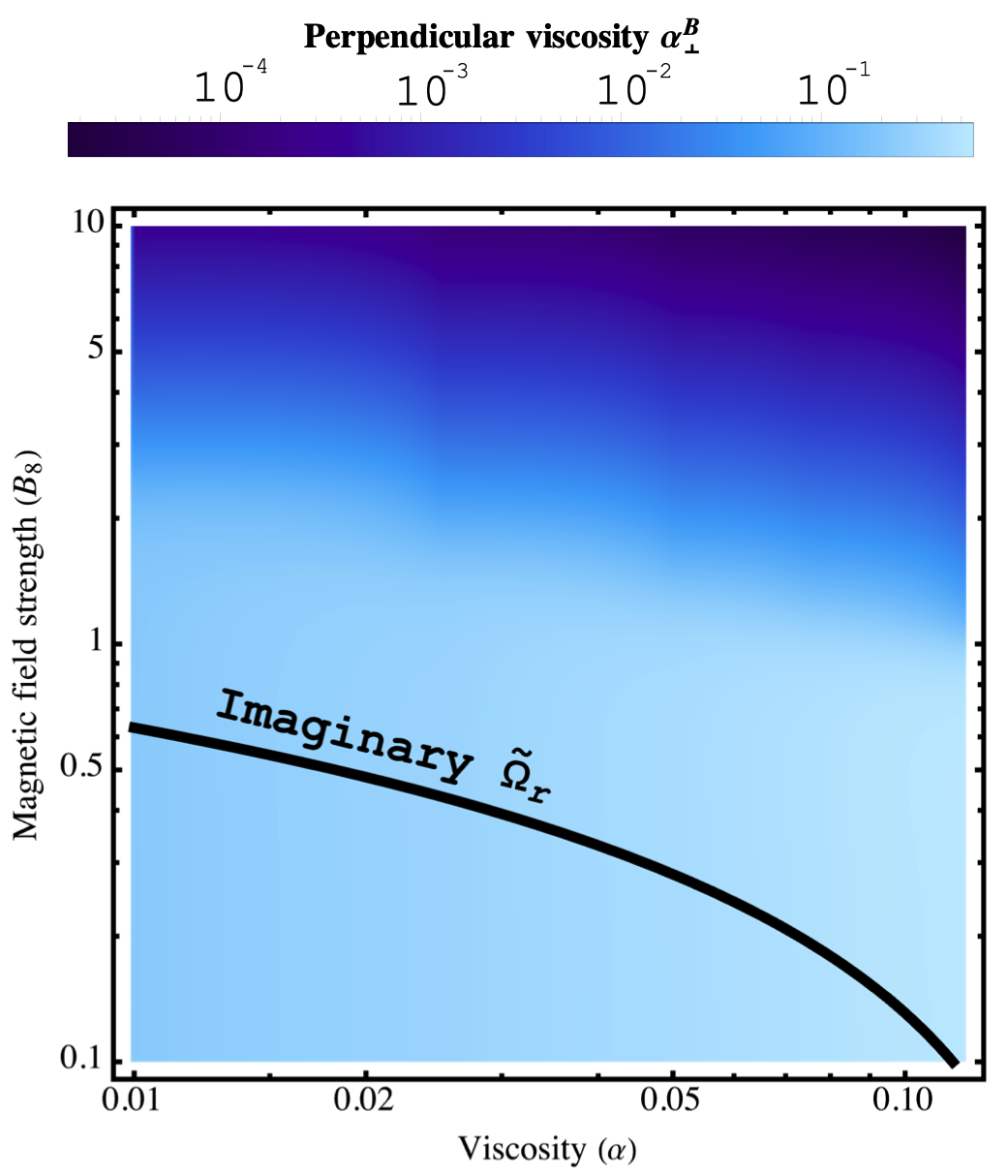}
 \caption{Similar to Fig.~\ref{fig:perp_viscosity1} but in the $B_{0}$-$\alpha$ plane and for the \emph{unnormalised} perpendicular viscosity. Darker shades indicate smaller values. 
 The region residing above in the solid curve corresponds to the regime where $\tilde{\Omega}_{r}^2 < 0$ and numerical treatments are strictly necessary as neither viscous nor 
 magnetic terms are negligible.}
 \label{fig:perp_viscosity}
\end{figure}

We can instead visualise the perpendicular viscosity in the $B_{0}$-$\alpha$ plane more directly, as in Fig.~\ref{fig:perp_viscosity}, showing unnormalised values. 
Darker shades are apparent at higher field strengths, indicating a \emph{reduction} as expected. This generally implies the disc is more prone to instabilities 
when strong fields are present as the vertical viscosity erodes. Moreover, we see a softer gradient in $\alpha_{\perp}^{\rm B}$ as we increase $\alpha$ prior to the 
transition curve where $\tilde{\Omega}_{r}$ becomes imaginary from expression \eqref{OmradB1}. In the neighbourhood of this curve, neither the inviscid or non-magnetic 
limits are valid: when $\tilde{\Omega}_{r} \sim 0$ we necessarily have $|\tilde{\Omega}_{r}|^2 \ll i \sigma \lambda$ within the dispersion relation \eqref{detomega}. 
Even beyond this point, as $\sigma$ is complex in general, one cannot na{\"i}vely discard viscous terms since the real and imaginary components do not exhibit a clear 
hierarchy. 

\subsection{The tearing radius revisited}
\label{sec:tearing2}

\begin{figure*}
\centering
 \includegraphics[width=0.96\textwidth]{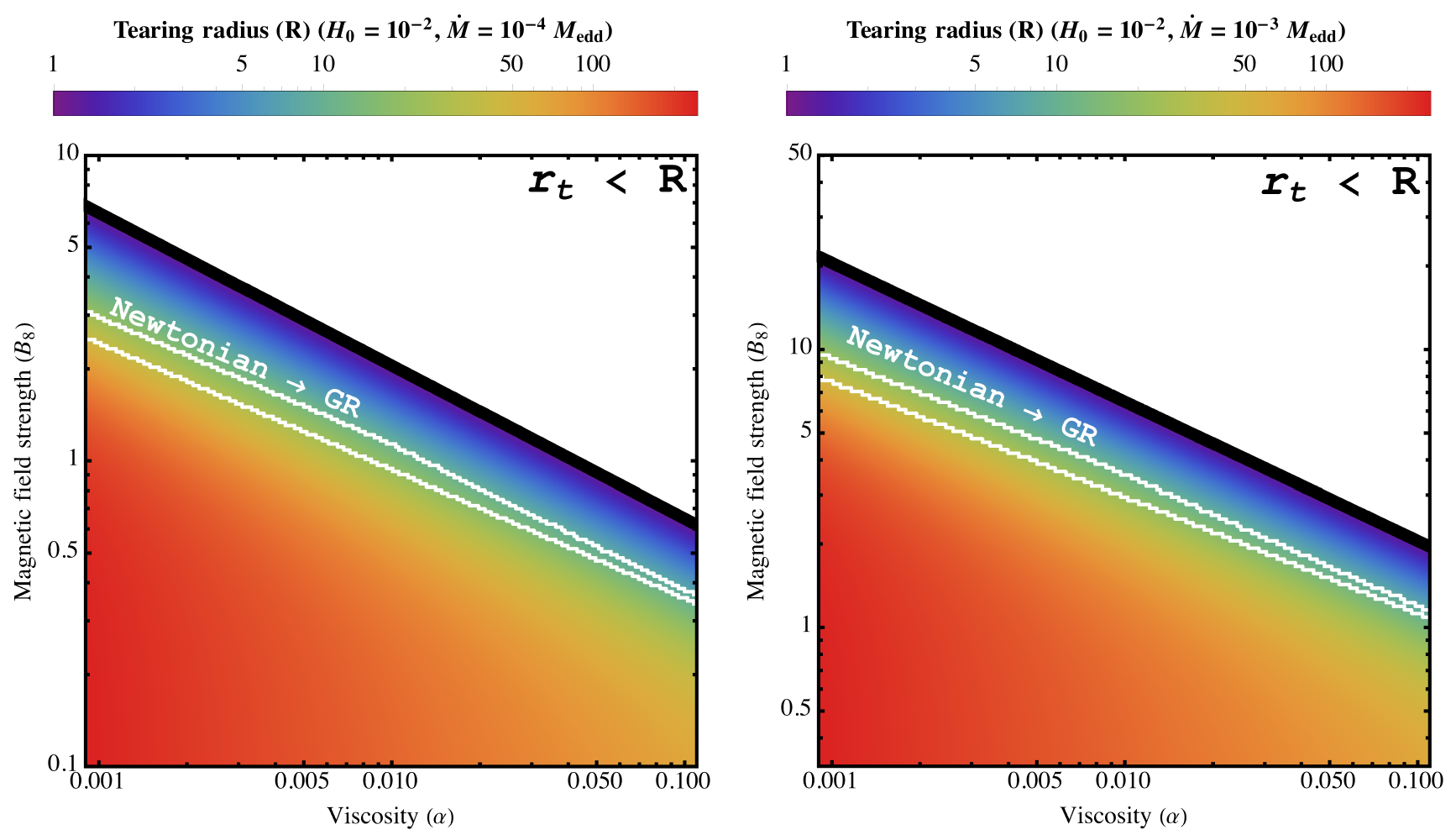}
 \caption{Tearing radii ($r_{t}$) in units of $R$, as a function of $B_{0}$ and $\alpha$, for fixed parameter choices $M = 1.4 M_{\odot}$, $R = 10$~km, 
 $H/r_0 = 10^{-2}$, $q=0.2$, and $\beta = |\psi| =  0.3$ for either $\dot{M} = 10^{-4} \dot{M}_{\rm edd}$ (left) or $\dot{M} = 10^{-3} \dot{M}_{\rm edd}$ (right). 
 Above the respective white contours (with the lower and upper lines delimiting the range $|\alpha_{\perp} - \alpha| > 20$ and $|\alpha_{\perp} - \alpha| > 10$, respectively) 
 one anticipates a transition to a relativistic regime, while above the black contour no solution exists as the tearing radius migrates beyond the stellar surface. 
 Redder shades indicate greater values of $r_{t}$. Note the different vertical axis scaling between the two panels.}
 \label{fig:rtwithB}
\end{figure*}

We are now finally in a position to evaluate the tearing radius \eqref{rteq3} with all ingredients included. One technical issue remains, which is that since 
$\gamma$ depends on the \emph{local} field strength (see Fig.~\ref{fig:gammafn}) we must construct a three-dimensional table $\alpha_{\perp}(\alpha,B_{0},r_{0})$. 
This is needed because we do not know which radius applies \emph{a priori}. This means that \eqref{rteq3} is no longer a simple cubic with closed-form solutions at 
equality since $\alpha_{\perp}$ varies significantly with $r_{0}$ through $B_{z}$. We use a brute-force strategy to find the roots: first solve Eqs.~\eqref{detomega} 
and \eqref{a2awithB1} simultaneously over a fine grid of $\alpha, B_{0}$, and $r_{0}$ values with an inbuilt flag that locates the avoided crossing by comparing 
neighbouring eigenfrequencies ($\tilde{\sigma}_{\rm R}$) and prints only values for $\alpha_{\perp}$ for the relevant branch. In practice, grids of size $96 \times 96 \times 52$
are used with log-spacing for $10^{-3} \leq \alpha \leq 0.1$, $10^{7} \lesssim B_{0}/\text{G} \lesssim 10^{10}$ and $R \leq r_{0} \lesssim 200 R$. The full relation for 
$\alpha_{\perp}$ is constructed through linear interpolation, and then inserted into \eqref{rteq3} to locate $r_{t}$ via a Newton-Raphson method starting from the Newtonian 
guess (available as an analytic root when $\alpha_{\perp}$ is independent of radius).

Two such examples of carrying about the above are depicted in Fig.~\ref{fig:rtwithB} with fixed values $M = 1.4 M_{\odot}$, $R = 10^{6}$~cm, $H/r_0 = 10^{-2}$,  
$\beta = |\psi_0| = 0.3$, $q = 0.2$ but different accretion rates. The white contours within the panels depict the equalities $|\alpha_{\perp} - \alpha| = 20 x$ (lower) 
or $|\alpha_{\perp} - \alpha| = 10 x$ (upper), roughly delimiting the validity of the Newtonian regime: above this band one anticipates a transition towards intermediate 
cases where relativistic effects become important. As expected, this also corresponds to tearing radii close to the star  ($\gtrsim$~few stellar radii) where magnetic 
pressures are prominent. 

That is, GR and magnetic effects tend to either be both important or unimportant, at least as far as how the $\alpha_{\perp}(\alpha)$ affects $r_{t}$ is concerned. For higher accretion rates (right panel), we find an upward 
shift in the sense that greater values of the field strength and/or viscosity parameter are required to obtain the same tearing radius. 
This occurs as the magnetic torque~\eqref{eq:magtorque} scales inversely with $\dot{M}$. Though not shown, disc thickness has a similar effect, in the sense that 
greater values of $H_{0} = H/r$ lead to smaller tearing radii. 

In any given model, as the field strength and viscosity increase, eventually the tearing radius reaches the 
stellar surface indicating no solution exists (shown by the thick, black line). For $\dot{M} = 10^{-4} \dot{M}_{\rm edd}$, this occurs already for 
$B \lesssim 5 \times 10^{8}$~G at low viscosities ($\alpha \gtrsim 10^{-3}$). As anticipated from the ratio \eqref{eq:ratiort}, a visually-indistinguishable figure is 
obtained if we were to use the non-magnetic $\alpha_{\perp}(\alpha)$ relation except at very low horizontal viscosities, $\alpha$. In any case, a wide range of 
tearing radii $1 \lesssim r_{t}/R \lesssim 10^{3}$ could be accommodated in LMXB-like systems. 
Typically, the anticipated value reduces as the field increases (again highlighting 
the destabilising effect of central-body-hosted Lorentz forces).

\section{Astrophysical connections} 
\label{sec:astro}

In the preceding sections we examined the structure and instability thresholds for warped discs in cases where magnetic fields and LT torques are important. 
In this section, we apply the relevant expressions to some astrophysical systems. For instance, LMXBs host primaries which are rapidly rotating but weakly 
(relative to the neutron star population) magnetised (Sec.~\ref{sec:lmxbs}), while pulsations and cyclotron-line measurements from some ultra-luminous 
X-ray sources (ULXs) indicate they are slowly rotating but strongly magnetised  (Sec.~\ref{sec:ulxs}). Before moving to particular systems though, we outline 
some general considerations with respect to anticipated values of the \citet{ss73} parameter (Sec.~\ref{sec:alphavalues}) and the dynamics of precessing 
rings (Sec.~\ref{sec:generalconsiderations}).

\subsection{Anticipated values of $\alpha$} 
\label{sec:alphavalues}

A review from both the theoretical and observational points of view for what values $\alpha$ ought to take can be found in \citet{king07}. 
The basic conclusion reached there is that observations indicate typical values of $\alpha \sim \mathcal{O}(0.1)$, while numerical simulations find 
much smaller values (as low as $\alpha \sim$~0.004 in the 3D protoplanetary disc study of Ref.~\cite{fn06}, for instance).

Recent observational data from discs around compact objects suggest somewhat lower values for $\alpha$. For LMXBs like Aql X-1 where the disc
 is illuminated by X-ray irradiation from inner regions, for instance, \citet{dub01} found that choosing $\alpha_{\rm h} \sim 0.2$ in `hot' (where hydrogen is ionized) 
 or $\alpha_{\rm c} \sim 0.03$ in `cold' regions best reproduces the recurrence time of X-ray transients and associated light curves. 
As there are uncertainties pertaining to the conversion efficiency of rest-mass energy into radiation and other systematics (e.g. hydrogen column density), 
different fits can be found in the literature. Recently, \citet{cob24} argued that $\alpha_{\rm h} \sim 0.1$ together with $\alpha_{\rm c} \sim 0.0075$ better 
match quiescent-state data for Aql X-1, with an even-lower $\alpha_{\rm c} \sim 0.005$ during the source's outburst phase in 2010. 

Values in the range $10^{-3} \lesssim \alpha \lesssim 10^{-1}$ could thus be considered `canonical' for thin discs around neutron stars. 
In systems accreting in the neighbourhood of the Eddington rate, comparable values of $\alpha \sim 10^{-2}$ were estimated by \citet{star04} using data 
from  the Palomar-Green quasar sample (see their table 2) and \citet{ham20} for the hyperluminous X-ray source HLX-1.

By contrast, advances in GRMHD modelling, which have permitted high-resolution capturing of shocks and the magnetorotational instability (MRI), 
skew towards \emph{much larger} values: Refs.~\cite{lis21,lis23} found $\alpha_{\rm eff} \sim \mathcal{O}(10)$ in the inner regions of highly-tilted discs. 
Comparative values were recently found using a different numerical code even for relatively low tilts ($\beta \lesssim 30^{\circ}$) \cite{curd25}, 
which notably increases further for rapidly-rotating primaries (see their figure 7). These large values were at least partially attributed to the fact that warp-related 
shock dissipation -- features that were difficult to capture in earlier studies -- leads to huge increases in particle temperature and hence the effective viscosity. 
Lower values of $\alpha$ are found at larger radii where the disc has not been radiatively-thickened (see, e.g., figure 10 in Ref.~\cite{lis23}). In particular, in regions 
where the disc fragments due to thermal or hydrodynamic instabilities and the density plummets, values as low as $\alpha_{\rm eff} \sim 10^{-2}$ have been found 
\cite{lis21}. It should be noted that estimates of $\alpha$ in numerical studies are subject to some degree of definitional variation, as one must select a time 
band and average the velocity components to wash out oscillatory motions.  

While a thorough survey of the literature lies beyond the scope of this paper, it is clear that there is some mismatch between the numerical and observational sectors. 
Importantly, the \citet{ss73} parameter is not itself a direct measure of the isotropic viscosity appearing within Navier-Stokes equations, and thus it is not obvious that 
an \emph{a priori} fixed functional form of the viscous stress-tensor ought to apply in astrophysical systems. This, together with the absence of shock modelling, could 
be (partially) responsible for such discrepancy between numerical and (semi-)analytic models. 

The upshot is that in cold discs where \emph{tearing} occurs -- relevant  for this paper -- values of $\alpha_{\rm eff} \sim \mathcal{O}(10^{-2})$ 
seem to be agreed upon in both the observational and theoretical sectors.

\subsection{Masses and timescales for torn rings} 
\label{sec:generalconsiderations}

Using the thin disc density profile~\eqref{eq:density1}, the amount of mass, $\Delta M$, stored within some annulus $r_{\rm in} \leq r \leq r_{\rm out}$ 
is easily estimated as
\be
\Delta M \approx 2 \pi \int^{r_{\rm out}}_{r_{\rm in}} dr \int^{H(r)}_{0} dz \, r \rho(r) 
 =  \frac{4 \dot{M} \left( r_{\rm out}^{3/2} - r_{\rm in}^{3/2} \right)} {9 H_{0}^2 \sqrt{G M} \alpha}.
 \label{eq:storedmass}
\ee
If disc tearing occurs, we can identify $\Delta M$ with the mass stored within some precessing ring that eventually crashes into a neighbour and subsequently 
donates its material to the compact object. As discussed by \citet{raj21b}, the timescale over which angular momentum cancellation may be expected is the 
differential nodal precession time
\be
\tau_{\rm dnp} = \frac{1}{\Omega_{\rm LT}(r) - \Omega_{\rm LT}(r + \Delta r)}
\approx \frac{G M}{6c^3} \frac{r}{\Delta r} \frac{\bar{r}^3}{q} ,
 \label{eq:dnp}
 \ee
where $\Delta r$ is the separation between the torn ring and its neighbours. In particular, it is reasonable to suppose that $\Delta r \approx H$ for strongly unstable rings, 
though it could plausibly be somewhat larger or smaller since the tearing process itself adjusts the hydromagnetic structure (e.g., the viscosity $\alpha$; see Ref.~\cite{lis21}). 
For a neutron star rotating with $q = 0.2$ (or $\nu \sim 400$~Hz), we estimate
\be \label{eq:tuffdnp}
\hspace{-0.2cm} \tau_{\rm dnp} \approx 0.02 \text{ yr} \times \left( \frac{r}{10^8 \text{ cm}} \right)^{3} 
\left(\frac{0.2}{q}\right) \left(\frac{10^{-3}}{H_0}\right) \left(\frac{ 1.4 M_{\odot}}{M}\right)^{2}.
\ee
The further away the ring is from the star the longer the precession time, expected since the LT frequency decays with radius. 
On the other hand, the mass stored~\eqref{eq:storedmass} also increases because the disc is thicker at larger radii (for constant $H/r$). 
Once the angular-momentum has been effectively cancelled it takes a further free-fall time for the material to land onto the stellar surface, 
though this is short relative to $\tau_{\rm dnp}$. 

As a lower limit (in the context of accreting, magnetised neutron stars) we may expect that the inner radius is set by the Alfv{\'e}n 
radius from expression \eqref{eq:alfven},
\begin{align}
\frac{R_{\rm A}}{R} &\approx 4.7 \left(\frac{\xi}{0.1}\right) \left( \frac{B_0}{10^9 \text{ G}} \right)^{4/{7}} \left( \frac{R}{10 \text{ km}} \right)^{{3}/{7}} 
\nonumber \\
&\quad \times \left( \frac{10^{-4} \dot{M}_{\rm edd}}{\dot{M}} \right)^{{2}/{7}} \left( \frac{1.4 M_{\odot}}{M} \right)^{{1}/{7}}.
\label{RAeq1}
\end{align}
The luminosity produced by the infalling ring mass can be estimated as
\be
L_{\rm ring} \sim \frac{G M}{R} \frac{  \Delta M}{ \tau_{\rm dnp}}.
\ee
For a system with a `background' luminosity (mostly emitted in the X-band for compact objects)
\be
L_{\rm X} \sim \frac{G M \dot{M}}{R},
\label{LX}
\ee 
this represents a fractional luminosity increase of
\be
\cL = \frac{L_{\rm ring}}{L_{\rm X}} \sim \frac{ \Delta M}{\dot{M} \tau_{\rm dnp}}.
\label{Lfrac}
\ee 
The ring's outer radius may be set by the (magnetically adjusted) viscous tearing radius from Sec.~\ref{sec:Btorque}, namely Eq.~\eqref{rteq3}. 
Typical values of this tearing radius are shown in Fig.~\ref{fig:rtwithB}. 

We now turn to estimating the relevant tearing masses \eqref{eq:storedmass} and torn-material accretion timescale \eqref{eq:tuffdnp} for several different 
astrophysical systems in the following sections.

\subsection{Low-mass X-ray binaries} 
\label{sec:lmxbs}

Even the most stable LMXBs containing a neutron star exhibit some degree of variability \cite{bahdeg23}. For example, observations of 
4U 1543--624 taken in 2020 reveal that the 0.5--50 keV luminosity was a factor $\gtrsim 2$ weaker than of the peak luminosity during its 2017 
brightening \cite{lud21}, while 4U 1705--44 \cite{mun02} and MAXI J0556--332 \cite{page22} have shown intermittent flux increases by factors 
of order $\sim 10$ from minimum to maximum over multi-year observational campaigns. Although the origin of large-amplitude X-ray brightenings 
is not fully understood, such episodes are likely attributed to enhanced mass-accretion rates \cite{ing13}. Such episodes may be triggered 
by dynamical instabilities taking place in the disc owing to phenomena associated with the host star(s), such as from non-planar tides \cite{lar96,dogan15}, 
radiation-reaction due to illumination \cite{pringle96,wij99}, or tearing \cite{lod10,nk12}. We consider the latter, especially supported in systems with `outburst' 
recurrence times that show no evidence of periodicity \cite{phil18}. 

As an example application of the formulae arrived at in Sec.~\ref{sec:magfield}, we focus on one of the more well-studied LMXBs, SAX J1808.4--3658 \cite{wij98}. 
The object has an inferred field strength of order $B_0 \sim 2 \times 10^{8}$~G with a relatively low quiescent X-ray luminosity, $L_{\rm X} \gtrsim 10^{32}$~erg/s \cite{hart09}, 
similar to the parameters chosen in Fig.~\ref{fig:gammafn}. As inferred from the standard relation~\eqref{LX}, the system accretes (on average) at a rate well below the Eddington one, 
$\langle \dot{M} \rangle \sim 10^{-5} \dot{M}_{\rm edd}$. 
Moreover, the detection of a relativistically-broadened iron line from this source led \citet{pap09} to infer that the inner edge of the accretion disc lies at 
$\approx 4.4^{+1.8}_{-1.4}$ Schwarzschild radii. This matches the Alfv{\'e}n radius from expression \eqref{RAeq1} provided we have $\xi \lesssim 0.1$ (see also Ref.~\cite{gs21}).

In 2015, the object went into outburst, with the X-ray flux increasing by a factor $\sim 60$ for $\sim$~10 days (see figure 2 in Ref.~\cite{sanna17}). If the brightening were triggered 
by an episode of enhanced accretion from disc tearing, it would imply the following two conditions:
\begin{equation}
 \label{eq:deltamcond}
\cL_{\rm J1808}  \sim 60, \qquad \tau^{\rm J1808}_{\rm dnp} \sim 10 \text{ days},
\end{equation}
which follow from~\eqref{Lfrac}. From expression \eqref{eq:tuffdnp}, we see that second condition within expression \eqref{eq:deltamcond} is satisfied provided 
the tearing radius is $r_{t} \approx 115 R$ (noting that SAX J1808.4--3658 has a spin frequency of $\nu = 400.98$~Hz).

\begin{figure}
\centering
 \includegraphics[width=0.48\textwidth]{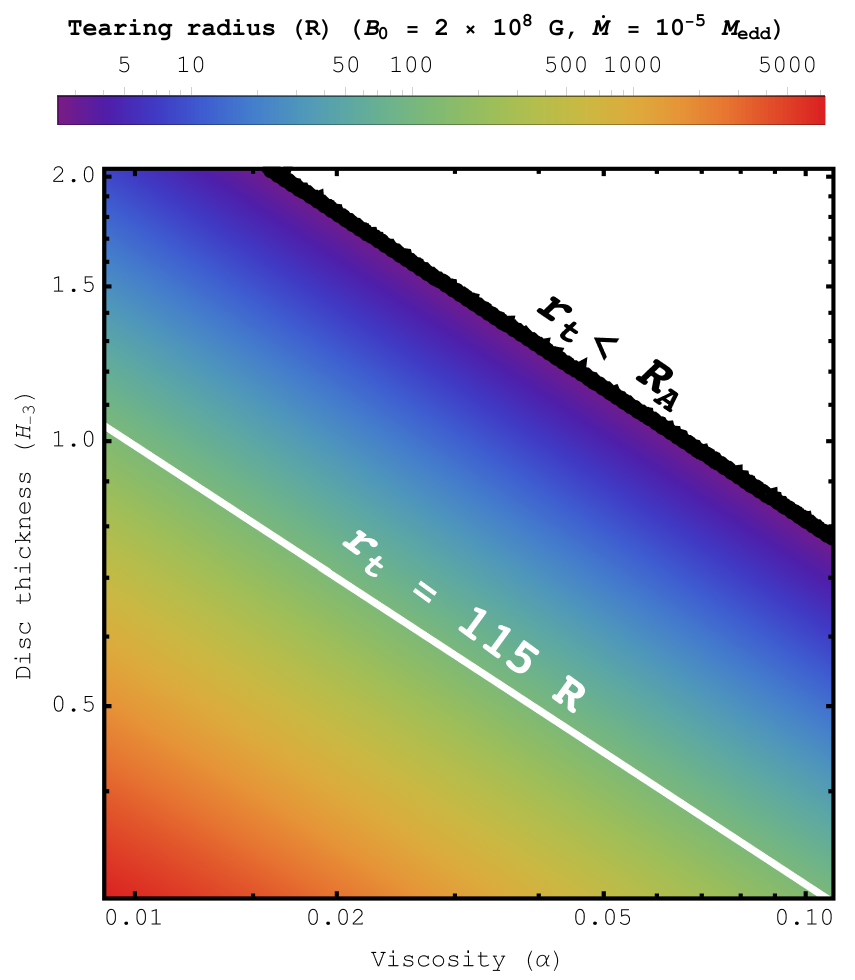}
 \caption{Tearing radius $r_{t}$ in units of the stellar radius for parameters appropriate to SAX J1808.4--3658 (see text), as a function of viscosity and disc 
 thickness (in units of $H_{-3} = 10^{3} \times H/r_0$). The white contour corresponds to $r_{t} = 115 R$, as anticipated from matching timescales within 
 expressions \eqref{eq:tuffdnp} and \eqref{eq:deltamcond}, with the black line representing the inner radius of the disc as determined by \protect\citet{pap09}.}
  \label{fig:SAXJ}
\end{figure}

Solving Eq.~\eqref{rteq3} in a manner similar to that carried out in Sec.~\ref{sec:local} but instead allowing the disc thickness and viscosity to vary, it is easy to show that
\begin{equation} \label{eq:h0exp}
H_{0}  \approx \frac{2 \sqrt{2} \left(\sin \beta\dot{M} q\right)^{1/2} (G M)^{3/4} }{c r_{t}^{1/4} 
   \left(3 \alpha  B_{0}^2 R^3+4 \alpha_{\perp} c \dot{M} r_{t} |\psi| \right)^{1/2}},
\end{equation}
for given $r_{t} = r_{\rm out}$. This implicit relation (noting that $\alpha_{\perp}$ depends on $H_{0}$ in the general, magnetoviscous case) is depicted in Fig.~\ref{fig:SAXJ} 
for $\beta = |\psi| = 0.3$ as a demonstration. Fixing $\dot{M}$, $B_{0}$, and $r_{\rm in}$ from the X-ray observations referenced above, $r_{t}$ from the inferred precessional 
timescale in \eqref{eq:deltamcond}, and $H_{0}$ from expression \eqref{eq:h0exp} through the tearing-radius equation, all that remains free in the estimate for $\Delta M$ 
are the alpha parameters, the inclination, and the warp. However, unless $\alpha_{\perp}$ is of order unity for these parameters, it is the magnetic terms that are most relevant 
(see expression~\ref{eq:ratiort}). In this limit, $\alpha$ drops out completely in the expression for $\Delta M$ and we find (noting $\sin \beta \sim \beta $ for small inclinations)
\begin{equation} \label{eq:deltamsax}
\Delta M_{\rm J1808} \approx 2 \times 10^{-13} \left(\frac{\beta}{0.03}\right)^{-1} M_{\odot} .
\end{equation}
While expression \eqref{eq:deltamsax} may appear small, given the low quiescent accretion rate in SAX J1808.4--3658 we find that
\begin{equation} 
\label{eq:ratiosaxj}
\cL_{\rm J1808} \approx 47 \left(\frac{\beta}{0.03}\right)^{-1},
\end{equation}
which is comparable to the observed value \eqref{eq:deltamcond}. While not a direct measure of the \emph{spin-orbit} misalignment, such a value of $\beta$ 
is roughly consistent with the pulse-profile and hotspot modelling of \citet{ibr09}, who estimated the colatitude of the hotspot centroid to be between 4 and 10 degrees. 
If the neutron star mass and radius were larger than the `canonical' values -- which would decrease the inferred $\dot{M}$ in quiescence -- the estimate \eqref{eq:ratiosaxj} 
increases further. We may therefore conclude that sensible parameters may explain the 2015 event via disc tearing. 

Brighter events by contrast, such as that seen from Aql X-1 where the luminosity jumps by $\sim$~4 orders of magnitude over $\sim$~month-long timescales 
\cite{kocab25}, can only be explained for very low inclinations. This source has comparable quiescent X-ray luminosity and field strength estimates to SAX J1808, 
though spins faster ($\nu = 550.3$~Hz). We may thus expect $r_{\rm out} \sim \text{few} \times 10^{8}$~cm by matching timescales using Eq.~\eqref{eq:tuffdnp}, giving
\begin{equation} \label{eq:deltamsax2}
\cL_{\rm Aql} \sim 2 \times 10^{2} \left(\frac{\beta}{10^{-2}}\right)^{-1},
\end{equation}
which cannot easily reproduce enhancements to the quiescent $\dot{M}$ exceeding factors of $\sim 10^{3}$ for $\sim$~months except for $\beta \ll 10^{-2}$.
On the other hand, Aql X-1 exhibits a bimodal distribution of brightening episodes with approximately half being considerably dimmer \cite{camp13}. 
In particular, the so-called `short-low' category of events lasting $\sim$~days \cite{trigo18} are readily explained by the tearing model as the ratio $\cL$
is large. 

We close this section by remarking that given the largeness of the $r_{t}$ values required to explain dramatic outbursts, the impact of magnetic and GR 
augmentations of the $\alpha_{\perp}(\alpha)$ relation are less relevant than changes to the $r_{t}$ condition itself; see Eq.~\eqref{rteq3}. 
However, many sources -- most notably the class of transitional millisecond pulsars (tMSPs) -- show X-ray variability even in quiescence on timescales 
as short as $\lesssim$~minutes (see, e.g., Figure 2 in Ref.~\cite{bag23}). Such unsteady flows could theoretically be explained by magnetic pumping, where the 
Alfv{\'e}n radius oscillates due to the propagation of warp modes and triggers short-term variability (see Footnote 3). Modelling such flickering would certainly 
require the use of the adjusted viscosity relation since we anticipate $r_{t} \sim R_{\rm A}$ from expression \eqref{eq:tuffdnp} (see also Fig.~\ref{fig:perp_viscosity}). 
Applications of the model developed here for the tMSP subclass of LMXBs will be explored elsewhere.

\subsection{Ultra-luminous X-ray binaries} \label{sec:ulxs}

ULXs are an enigmatic class of exceptionally bright systems. While a number of these sources contain black holes, some of them show coherent 
pulsations  -- the so-called PULXs -- and must therefore host neutron star primaries. As the name implies, these sources all show $\dot{M} > \dot{M}_{\rm edd}$ 
in outburst but also spin slowly ($q \sim 10^{-4}$).  PULXs thus reside in a different parameter space to LMXBs, altering the dynamical picture of the magnetoviscous disc 
and any would-be tearing.

While there is some debate in the literature, the expectation is that PULXs host magnetic field strengths of at least $B_0 \gtrsim 10^{12}$~G. 
For instance, a $\sim 130$~keV cyclotron resonance feature was observed in the spectrum of the Galactic PULX Swift J0243.6+6124, implying a magnetic field 
strength of $\gtrsim 10^{13}$~G near the surface \cite{kong22}. 

Due to the combination of a strong field, large $\dot{M}$, and small $q$, we expect that disc tearing 
is unlikely to play a significant role in these system. This is because the tearing radius \eqref{rteq3} dips within the Alfv{\'e}n sphere for realistic values of the 
disc thickness and viscosity, unless the field is dominated by higher multipoles due to magnetic burial or otherwise \cite{suvm19,suvm20,gs21}. Moreover, since the mass 
accretion rate is so high, one expects that the disc will be thickened by radiation pressures \cite{ws88,chash17} and thus the thin-disc model employed here 
may not apply in general.

Nevertheless, X-ray pulsations were discovered from Swift J0243.6+6124 even at luminosities of $L_{\rm X} \sim 3 \times 10^{34}$~erg/s \cite{dor18}, 
indicating a quiescent accretion rate as low as $\langle \dot{M} \rangle \sim 10^{-4} \dot{M}_{\rm edd}$. Between 2017 and 2018, the object underwent 
a `giant outburst' where the luminosity reached super-Eddington levels ($L_{\rm X} \sim 2 \times 10^{39}$~erg/s \cite{wil18,tao19}). 
If we were to interpret this event via disc tearing we may write, similar to Eq.~\eqref{eq:deltamcond},
\begin{equation} 
\label{eq:deltamcond0234}
\cL_{\rm J0243} \sim 10^{5}, \qquad \tau^{\rm J0243}_{\rm dnp} \sim 1 \text{ yr}.
\end{equation}
Matching such values to the theoretical estimations is difficult even for extreme parameter choices. Fixing $q = 10^{-4}$, we get from Eq.~\eqref{eq:tuffdnp} 
that
\begin{equation} 
\label{eq:rout0234}
r_{\rm out} \sim 3 \times 10^{7} \left( H_{0} / 10^{-3} \right)^{1/3} \text{ cm},
\end{equation}
in order to match timescales. Further taking a conservative $B_0 = 10^{12}$~G for the dipole component and $\xi = 10^{-2}$ to explain a compact 
magnetosphere \cite{dor18}, matching expression \eqref{eq:storedmass} to that within \eqref{eq:deltamcond0234} implies a tight relationship between 
$H_{0}$ and $\alpha$. If $H_{0} \approx 10^{-3}$ in line with standard assumptions, for instance, this demand amounts to an impossibly small $\alpha \sim 10^{-12}$.

Even ignoring the above, the tearing-radius solution to equation \eqref{rteq3} cannot accommodate 
expression \eqref{eq:rout0234} for any reasonable range of $H_{0}$ and $\alpha$. The problem becomes more severe if one uses larger values of $B_{0}$, such as that 
inferred from quiescent spindown data \cite{serim23}, even if accounting for GR increases to the spindown luminosity \cite{ssp25}. For the less dramatic outburst 
from M82 X-2 around June 2015, where $L_{\rm X}$ was larger by a factor $\gtrsim 10^{2}$ compared to its quiescent return in June 2016 \cite{bach20}, it is still 
difficult to match data unlike in Aql X-1. 

We conclude therefore that brightening episodes in PULXs are unlikely to be explained by tearing or that extra ingredients (e.g., a thick disc) 
are needed on the modelling side. 

\subsection{Active galactic nuclei} \label{sec:agn}

{The noncolinearity of high-velocity declinations in astrophysical water masers operating in several AGN (see, e.g., Ref.~\cite{hern96}) indicate that discs in such environments are often warped.
While there are many subtleties in reality, AGN can be broadly separated into two categories: those that show radio-loud jets and those that do not (or only weakly or intermittently).
It remains a largely open problem as to exactly \emph{why} there is such a dichotomy, though linear polarisation \cite{park22} and rotation measure \cite{asada02} data support the position that radio-loud AGN are considerably more magnetised than radio-quiet ones. }

{In the standard picture where $\alpha_{2} \propto \alpha^{-1}$, the vertical shear viscosity is more efficient than the horizontal one ($\nu_{\perp} \gg \nu_{||}$). 
This means that the warp propagation timescale is much shorter than the time required for the global surface density to adjust under the influence of azimuthal shear viscosity (i.e., $t_{\rm warp} \ll t_R$ in the notation of Ref.~\cite{natar98}).
Such a scale separation, and disc-spin alignment more generally, is what is thought to keep astrophysical jets directionally stable \cite{natar99}.
In this picture, strong magnetic fields are the discriminating ingredient simply because the system has an easier time collimating a jet if in a MAD state (see Ref.~\cite{mck13} for a thorough discussion). }

{In the cases studied here with weak magnetic fields (`SANE' regime), we find a comparatively small perpendicular viscosity (Eq.~\ref{eq:newtampformula}) compared to the standard hydrodynamical case ($\alpha_{\perp} \approx 1/2 \alpha$ \cite{og99}). 
If such a `revision' applies, we instead have $t_{\rm warp} \lesssim t_R$ rather than an obvious hierarchy (compare curves in Fig.~\ref{fig:differentalpha}).
Without an efficient perpendicular viscosity to smooth the inner disk into alignment, global warps may be able to persist over timescales comparable to accretion.
Because jets are thought to be launched perpendicular to the innermost disc plane, we expect the jet axis to wander and precess over long timescales if $t_{\rm warp} \sim t_R$ (see Ref.~\cite{cui23} for a discussion on observations of the jet nozzle in M87, apparently precessing with a period of $\sim$~11 years).}

{By contrast, in the magnetic model considered here with a large plasma-beta (i.e., MAD disks), we argue that $\alpha_{\perp} \propto \alpha^{-3/2}$ (Fig.~\ref{fig:perp_viscosity1}).
The system is thus naturally in a $\nu_{\perp} \gg \nu_{||}$ regime, and we expect again that jets can form because of timescale hierarchies.
In this respect, revisions to the perpendicular viscosity fit naturally with the dichotomy of radio-loud AGN: those with strong fields are such that $t_{\rm warp} \ll t_R$ indicating stability, while those without have $t_{\rm warp} \lesssim t_R$ and launching jets becomes difficult as there is no stable platform.
It could thus be that, although the magnetic field is implicitly responsible for changing the viscosity profile, it is ultimately just the relationship $\alpha_{\perp}/\alpha$ that controls whether jets can reliably launch.}

{One avenue for falsifying the model -- beyond the scope of this paper to investigate thoroughly -- is that of a correlation between precessional motion of the jet and the magnetic field strength (as we predict $\alpha_{\perp} \propto B^{-4}$).
In other words, if the field is \emph{too} strong, we could again return to a case of $t_{\rm warp} \sim t_R$ despite the $\alpha_{\perp} \propto \alpha^{-3/2}$ scaling if the plasma-beta $\gg 1$ (see the top panel of Fig.~\ref{fig:perp_viscosity1}).
}

{Another interesting avenue concerns the elusive \emph{quasi-periodic eruptions} (QPEs) \cite{min19}. 
A promising model of these events involves an unstable disc, where a strong magnetic field is invoked to fast-track the instability that leads to an eruption \cite{kaur23}.
Since we find that warp instabilities set in precisely around the magnetically-dominated regime (see Fig.~\ref{fig:instab}), discs in such a state could naturally experience 
tearing without any external (e.g. tidal) influence.
Given the reduction of $\alpha_{\perp}$ as a function of $B_{0}$ (Tab.~\ref{tab:verticalvisc}), it would be worth revisiting the viability of that scenario since we anticipate 
that the optical depth -- controlling runaway heating events \cite{pan22} -- scales inversely with the viscosity \cite{ss73}.}

\begin{table*}
\centering
\caption{Summary of key scaling relations between the vertical viscosity, $\alpha_{\perp}$, to the horizontal viscosity, $\alpha \ll 1$, and other 
parameters (as defined in text) in a thin, warped disc.}
\hspace{-1.2cm}\begin{tabular}{ccc}
 \hline
 \hline
Model & Relation & Reference \\
\hline
Newtonian, local (undamped) & $\alpha_{\perp} = 1/\alpha$ & \citet{lp07} \\
Newtonian, hydrodynamical & $\alpha_{\perp} \approx 1/ 2 \alpha$ & \citet{pp83,og99} \\
Newtonian, local (damped) & $\alpha_{\perp} \approx \alpha/(\alpha_{\perp} - \alpha)^{2}$ & This work; Eq.~\eqref{alphaperpN} \\
GR-dominated, local & $\alpha_{\perp} \approx \alpha /x^2$ & This work ($x^{3} \ll \alpha \ll 1$); Eq.~\eqref{alphaperpGR} \\
GR, local & Numerical from Eqs.~\eqref{eq:hydroamp} and \eqref{perpviscnewt} & This work ($B_{0} = 0$); Fig.~\ref{fig:hydro_viscosities} \\
Local, magnetically-dominated & $\alpha_{\perp} \propto \alpha^{-3/2} B_{0}^{-4}$ & This work ($\gamma r_{0}^2/H^2 \gg 1$); Eq.~\eqref{eq:magviscreln} 
and Fig.~\ref{fig:perp_viscosity1} \\ GR, local, magnetoviscous & Numerical from Eqs.~\eqref{detomega} and \eqref{a2awithB1} & This work; Fig.~\ref{fig:perp_viscosity} \\
\hline
\hline
\end{tabular} 
\label{tab:verticalvisc}
\end{table*}

{As another example pertinent to AGN, data from the James Webb Space Telescope (JWST) are revealing a vast population of high-redshift galaxies that are extremely 
massive \cite{haro23,lamb24}. 
Their abundance poses a challenge for practically all cosmological evolution scenarios \cite{luca25}; put simply, how is it possible 
for the black holes at the heart of these structures to grow so fast? 
If evidence is found for even moderately-strong magnetic fields in these systems, episodes of hyper-accretion would be easier to instigate due to magnetic-adjustments in the tearing dynamics described in Sec.~\ref{sec:tearing2}.}

\section{Discussion}
\label{sec:discussion}

In this paper, we reexamine key assumptions considered in the literature used to estimate the perpendicular viscosity coefficient, $\alpha_{\perp}$, 
in warped accretion discs and associated instability thresholds (Sec.~\ref{sec:local}). In addition, we incorporate GR effects and externally-sourced Lorentz forces
 at the level of a local fluid analysis (Sec.~\ref{sec:magfield}) semi-analytically for the first time (though see also Refs.~\cite{po18,lis21,lis23}). 
Depending on the type of compact binary, either or both of these effects may be important and we find that the $\alpha_{\perp}$--$\alpha$ relationship 
can change dramatically; a summary of viscosity relations is given in Table~\ref{tab:verticalvisc}. 
When the field becomes sufficiently intense (quantified by expression \ref{eq:gammaeqn0}), we find that the classical warp modes transition 
into a magnetically-dominated family exhibiting avoided crossings (Fig.~\ref{fig:avoided}). 
On the physical side, the fact that $\alpha_{\perp}$ tends to decrease as a function of magnetic field strength (Figs.~\ref{fig:perp_viscosity1} 
and \ref{fig:perp_viscosity}) implies that \emph{tearing} is easier to achieve in general (quantified by Eq.~\ref{rteq3} and Figs.~\ref{fig:rtwithB} and \ref{fig:SAXJ}). 
For compact binaries involving neutron stars, for example, this means that episodes of rapid mass infall triggered by warp instabilities appear as an 
even stronger candidate to explain outburst transitions (Sec.~\ref{sec:lmxbs}). 

As in all studies of accreting compact objects, there are many parameters to consider. 
Most of these are unknown in general since they relate to microphysical interactions (e.g., the base viscosity $\alpha$) or macrophysical orientation 
(e.g., the tilt angle $\beta$; see Fig.~\ref{fig:schematic}).  With respect to the latter, recent polarimetric instruments such as the Imaging X-ray Polarimetry 
Explorer (IXPE) are improving the prospects for estimating inclination angles in neutron-star X-ray binaries \cite{doro22,kash25}.
These data will be useful in future to more accurately assess the likelihood of tearing as an explanation for brightening episodes and hard-to-soft state transitions. 
The observation of linear X-ray polarisations at the level of a few percent in sources like Cygnus X--1 \cite{kraw22} could be interpreted via an edge-on viewing of 
a warped disc \cite{ing15}, suggesting perhaps that tilt is common in Nature.

{Aside from the systems considered in Sec.~\ref{sec:astro}}, another case where the results obtained here may be relevant concerns the evolution of the recently-discovered `long-period transients' (LPTs) \cite{hw22}. 
These sources pulse in the radio band in a manner remarkably similar to radio-loud magnetars, but with periods ranging from several minutes to hours. 
While their origin is unknown, the two leading models involve `spun-down' magnetars \cite{cz24,suvm23} or compact white-dwarf binaries \cite{qz25,sdp25b}. 
In either case, supernova fallback or disc-fed accretion from a companion could play a role.
Given that strong magnetic fields are required to produce coherent radio emissions, any would-be discs near such systems could be in the magnetically-dominated regime.
In the binary picture, for example, the low duty cycle of many sources (such as GLEAM-X J 162759.5--523504.3 \cite{hw22}) could theoretically be explained by disc 
recession: warp-related instabilities deplete chunks of material such that magnetospheric interactions, otherwise impeded by the circling plasma, can take place to launch radio pulses.
Such events could be sporadic depending on the refilling time, explaining LPT irregularity. 
It would therefore be interesting in future to consider the tearing timescales described in Sec.~\ref{sec:astro} but for cataclysmic variables or polars rather than neutron
 stars or black holes. 

\section*{Acknowledgements}
We thank Suzan Do{\u{g}}an and Gordon Ogilvie for helpful correspondence on the nature of the $Q_{i}$ coefficients and their computational routines. 
AGS is grateful for support provided by the Conselleria d'Educaci{\'o}, Cultura, Universitats i Ocupaci{\'o} de la Generalitat Valenciana through Prometeo 
Project CIPROM/2022/13 during the early stages of this work, the European Union's Horizon MSCA-2022 research and innovation programme `EinsteinWaves' 
under grant agreement No. 101131233, the Deutsche Forschungsgemeinschaft individual research grant 570901071, and the High Performance and Cloud 
Computing Group at the Zentrum f{\"u}r Datenverarbeitung of the University of T{\"u}bingen (Project `AnqaGW').



%


\appendix

\section{The `exact' orbital oscillators}
\label{app:testbody}

This appendix provides a compact derivation of the equations of motion of a test body in a nearly circular/nearly equatorial  
orbit in the Kerr spacetime. 

The textbook geodesic equations are~\cite{bar72}:
\begin{align}
\nonumber
\Sigma^2 \dot{r}^2 &= V_r (r) 
\\
&= \left [  \cE ( r^2 + a^2 ) - a \cL \right ]^2 - \Delta \left [ r^2 + ( \cL - a \cE)^2 + \cQ \right ],
 \\
 \Sigma^2 \dot{\theta}^2 &= V_\theta (\theta) = \cQ - \cos^2\theta \left [ a^2 ( 1 -\cE^2 ) + \frac{\cL^2}{\sin^2 \theta} \right ],
 \\
 \Sigma \dot{t} & =  V_t =  \frac{r^2 + a^2}{\Delta} [  \cE ( r^2 + a^2 ) - a \cL] - a ( a \cE \sin^2\theta -\cL ),  
\end{align}
where $\Delta = r^2 - 2 M r + a^2$ and  $\Sigma= r^2 + a^2 \cos^2 \theta$. The constants $\{\cE, \cL, \cQ \}$  are per unit test body mass. 
We consider a nearly circular and nearly equatorial orbit, that is, the body's radial and meridional position as a function of its proper time $\tau$, 
is described as
\be
r(\tau) = r_0 + \delta r (\tau), \qquad \theta (\tau) = \frac{\pi}{2} + \delta \theta (\tau).
\ee
The constant energy and angular momentum, $\cE_0, \cL_0$ of the circular orbit are determined by 
\be
V_r (r_0) = 0 = V_r^\p (r_0).
\ee
(For the same orbit the Carter constant is $\cQ_0 = 0$.) Then, perturbing the above geodesic equations leads to the oscillator equations
\be
\delta \ddot{r} + \bar{\Omega}_r^2 \delta r= 0, \qquad  \delta \ddot{\theta} + \bar{\Omega}_\theta^2 \delta \theta= 0,
\ee
with
\be
\bar{ \Omega}_{r/\theta}^2 = - \frac{V_{r/\theta}^{\pp}}{2 \Sigma^2} \Big |_{r= r_0, \theta = \pi/2}. 
\ee
These are proper time frequencies; we can easily convert these expressions to coordinate time oscillators 
\be
\frac{d^2 \delta r}{dt^2} + \Omega_r^2 \delta r= 0, \qquad   \frac{ d^2 \delta \theta}{dt^2} + \Omega_\theta^2 \delta \theta= 0,
\ee
with
\be
\Omega_{r/\theta} = \frac{ \bar{ \Omega}_{r/\theta}}{\dot{t}} =   \frac{1}{\sqrt{2}V_t} \left ( - V_{r/\theta}^{\pp} \right )^{1/2} \Big |_{r= r_0, \theta = \pi/2}.
\ee
In terms of the dimensionless parameters $q =a/M, x = M/r_0$ we find
\begin{align}
\Omega_r^2 &= \frac{x^3 - 6 x^4 + 8 q x^{9/2} - 3 q^2 x^5}{(1+q x^{3/2} )^2},
\\ \nonumber \\
\Omega_\theta^2 &= \frac{x^3 - 4 q x^{9/2} + 3 q^2 x^5}{(1+q x^{3/2} )^2}.
\end{align}

\label{lastpage}

\end{document}